\documentclass[a4paper,11pt]{article}
\usepackage{jheppub} 
\usepackage{lineno}
\usepackage{xcolor}
\usepackage{extarrows}
\usepackage{braket}
\usepackage{enumerate}
\usepackage{caption}
\usepackage{subcaption}
\usepackage{booktabs}
\usepackage{float}
\usepackage{comment}
\usepackage{mathrsfs}

\newcommand{\mmod}{{\mathrm{mod}\ }}
\newcommand{\mA}{\mathrm{A}}
\newcommand{\mB}{\mathrm{B}}
\newcommand{\mC}{\mathrm{C}}
\newcommand{\ma}{\mathrm{a}}
\newcommand{\mb}{\mathrm{b}}
\newcommand{\mc}{\mathrm{c}}

\arxivnumber{} 

\title{Multi-entropy from Linking in Chern-Simons Theory}

\author[a]{Ma-Ke Yuan,}
\author[a]{Mingyi Li}
\author[a,b]{and Yang Zhou}

\affiliation[a]{Department of Physics and Center for Field Theory and Particle Physics, \\Fudan University, Shanghai 200433, China}
\affiliation[b]{Peng Huanwu Center for Fundamental Theory, Hefei, Anhui 230026, China}

\emailAdd{mkyuan19@fudan.edu.cn}
\emailAdd{limy22@m.fudan.edu.cn}
\emailAdd{yang\_zhou@fudan.edu.cn}

\abstract{
We study the multipartite entanglement structure of quantum states prepared by the Euclidean path integral over three-manifolds with multiple torus boundaries (the so-called link states) in both Abelian and non-Abelian Chern-Simons theories. For three-component link states in the Abelian theory, we derive an explicit formula for the R\'enyi multi-entropy in terms of linking numbers. We further show that the genuine multi-entropy faithfully quantifies the tripartite entanglement generated by GHZ-states, consistent with the fact that the prepared states are stabilizer states.}

\begin{document}
\noindent\makebox[\textwidth][r]{\texttt{USTC-ICTS/PCFT-25-49}}%
\vspace{-3em}
\maketitle
\flushbottom
\section{Introduction}
\label{sec:intro}

Understanding multipartite entanglement has become increasingly important in both quantum field theory and holography. Within the framework of the AdS/CFT correspondence, multipartite entanglement among boundary degrees of freedom is believed to underlie the emergence of bulk spacetime geometry. From the field-theoretic perspective, multipartite entanglement serves as a refined probe of nonlocal quantum correlations, topological order, and critical phenomena, extending beyond the information captured by bipartite entanglement entropy. Developing quantitative and conceptual tools to characterize multipartite entanglement is therefore crucial for elucidating the microscopic organization of quantum many-body systems and for advancing our understanding of holographic duality.

In Ref.~\cite{Gadde:2022cqi}, the multi-entropy was introduced as a natural multipartite generalization of the entanglement entropy. Its definition is based on the replica trick (see Section~\ref{subsec-multiE} for a brief review), where one computes Rényi-type quantities for integer replica number and performs an analytic continuation to $n = 1$. In general, this continuation is technically challenging and often inaccessible in quantum field theory. In this work, we demonstrate that for link states in Chern-Simons theory, the analytic continuation can be carried out explicitly, leading to a closed-form expression for the tripartite multi-entropy of the three-component link state~\eqref{eq-n-multiE-3link}.

Another central result of this work is the discovery of a network of relations among different bipartite and tripartite entanglement measures for link states. These include the entanglement entropy, mutual information, logarithmic negativity (and the third-order negativity), as well as the genuine multi-entropy. Through these relations, we uncover the precise physical interpretation of the tripartite genuine multi-entropy and clarify its role in characterizing multipartite correlations beyond pairwise entanglement. In particular we show that the genuine multi-entropy faithfully quantifies the tripartite entanglement generated by GHZ-states, consistent with the fact that the prepared states are stabilizer states.

Our results also fit naturally into a broader line of recent developments on multipartite entanglement and the Rényi multi-entropy. For instance, studies of multipartite correlation measures include Refs.~\cite{Walter:2016lgl, Nezami:2016zni, Zou:2020bly}; Refs.~\cite{Penington:2022dhr, Gadde:2024taa} discuss bulk replica symmetry breaking in the context of Rényi multi-entropy; Ref.~\cite{Gadde:2023zzj} provides a classification of multipartite entanglement; Refs.~\cite{Gadde:2023zni, Gadde:2024jfi} explore its monotonicity properties; Ref.~\cite{Harper:2024ker} performs field-theoretical computations in two-dimensional CFTs at small Rényi index; Refs.~\cite{Liu:2024ulq, Sheffer:2025jtc} study applications in topologically ordered systems; Ref.~\cite{Yuan:2024yfg} extends the construction to mixed states and Ref.~\cite{Iizuka:2025elr} further discusses the multipartite markov gaps; Refs.~\cite{Iizuka:2024pzm, Iizuka:2025ioc, Iizuka:2025caq} formulate the genuine version of multi-entropy and also investigate applications to black hole information problems; and Ref.~\cite{Harper:2025uui} employs Rényi multi-entropy to probe quantum critical points.

A particularly relevant line of research concerns link states in three-dimensional Chern-Simons theory, which provide concrete and topologically nontrivial realizations of multipartite entanglement. Entanglement entropy of link states has been analyzed in Refs.~\cite{Balasubramanian:2016sro, Dwivedi:2017rnj, Balasubramanian:2018por, Dwivedi:2020jyx}, while the relation between linking and quantum anomalies was explored in Refs.~\cite{Hung:2018rhg, Zhou:2019ezk}. Multipartite entanglement of link states was further investigated in Refs.~\cite{Salton:2016qpp, Balasubramanian:2025kaf}.

The remainder of this paper is organized as follows. In Section~\ref{sec-review}, we review the definitions and basic properties of the multi-entropy and the genuine multi-entropy. Section~\ref{sec-GME-links} is devoted to the study of genuine multi-entropy for link states in three-dimensional Chern-Simons theory, both in the Abelian and non-Abelian cases. In Section~\ref{sec-nega}, we compute the entanglement negativity for three-component links. 
In Section~\ref{sec-relations}, we clarify the physical meanings of genuine multi-entropy and logarithmic negativity for stabilizer states.
Additional proofs and technical details are collected in the appendices.

\paragraph{Note added:} While this paper was being completed, Ref.~\cite{Akella:2025owv} appeared on the arXiv, which has some overlap with our multi-entropy results. We have added Appendix~\ref{appx-stabilizerFormula} to clarify the relation.

\section{Review of (genuine) multi-entropy}\label{sec-review}
\subsection{Multi-entropy}\label{subsec-multiE}
Before introducing the multi-entropy, let us briefly review the concepts of entanglement entropy (EE) and R\'enyi entropy. Given a pure state $\ket{\psi}$ and dividing the entire system into $A$ and its complementary $\bar{A}$, the EE between them is defined as
\begin{equation}\label{eq-EE}
S(A) = - \text{Tr}\rho_{A} \log \rho_{A}\ ,
\end{equation}
where $\rho_{A} = \text{Tr}_{\bar{A}} \ket{\psi} \bra{\psi}$ is the reduced density matrix. 
R\'enyi entropy is a one-parameter generalization of EE
\begin{equation}\label{eq-RenyiE}
    S_{n}(A) = \frac{1}{1-n}\log\text{Tr}\rho_{A}^{n}\ ,
\end{equation}
and it reduces to EE~\eqref{eq-EE} in the limit $n \to 1$. This method of computing EE is known as the \textit{replica trick}. The R\'enyi entropy satisfies\footnote{In this paper we sometimes use the notation $S_n(A;\bar{A})$ to emphasize that it is the entanglement between $A$ and its complement $\bar{A}$. }
\begin{equation}
S_n(A) = S_n(\bar{A}) = S_n(A;\bar{A})\ .
\end{equation}
The $\mathrm{Tr}\rho_{A}^{n}$ part in the definition of R\'enyi entropy \eqref{eq-RenyiE} can be reformulated through the permutation and contraction of the indices of the $n$ copies of $\rho_{A}$, 
\begin{equation}
\begin{split}
\rho_{A}^{n} 
&= \left(\rho_A\right)_{\alpha_{1}}^{\alpha_{2}} \left(\rho_A\right)_{\alpha_{2}}^{\alpha_{3}} \left(\rho_A\right)_{\alpha_{3}}^{\alpha_{4}} \cdots \left(\rho_A\right)_{\alpha_{n}}^{\alpha_{1}}\\ 
&= \left(\rho_A\right)_{\alpha_{1}}^{\alpha_{\sigma\cdot 1}} \left(\rho_A\right)_{\alpha_{2}}^{\alpha_{\sigma\cdot 2}} \left(\rho_A\right)_{\alpha_{3}}^{\alpha_{\sigma\cdot 3}} \cdots \left(\rho_A\right)_{\alpha_{n}}^{\alpha_{\sigma\cdot n}}\ .
\end{split}
\end{equation}
with $\sigma = (123\cdots n)$ the permutation acting on the index of replicas.
Guided by this key observation, in \cite{Gadde:2022cqi} the authors defined the \textit{multi-entropy} $S^{(\mathtt{q})}$, which generalizes the concept of EE to $\mathtt{q}$-partite cases through the permutation and contraction of indices. Let us focus on tripartite pure state $\ket{\psi}_{ABC}$ to illustrate.

Like R\'enyi entropy, after preparing several copies for the tripartite system, the entanglement measure $S_{n}^{(\mathtt{q} = 3)}$ should be determined by the choice of permutation $(\sigma_{A},\sigma_{B},\sigma_{C})$ acting on the indices of different replicas, i.e., the way of contracting multiple density matrices. Note that there is an equivalence relation 
\begin{equation}
(\sigma_{A},\sigma_{B},\sigma_{C}) \sim (\sigma_{A},\sigma_{B},\sigma_{C})\cdot g
\end{equation}
allowing us to consider only $(\mathtt{q} - 1)$-parties, i.e., we can always choose $g=\sigma_{A}^{-1}$ to make $\sigma_{A} = \text{id}$ and the entanglement measure is only determined by $(\sigma_{B},\sigma_{C})$.\footnote{In this paper, we treat all the subsystems symmetrically. } Both $B$ and $C$ have $n$ replicas and $(\sigma_{B},\sigma_{C})$ are chosen to be cyclic permutations of order $n$.
In total, we have $n^{\mathtt{q} - 1} = n^2$ replicas, which form a $n \times n$ square lattice. The multi-entropy is defined by choosing the replica symmetry as $\mathbb{Z}_n \otimes \mathbb{Z}_n$ with the first/second $\mathbb{Z}_n$ acting on the subregion $B$/$C$ and cyclically permuting the replicas located in the same row/column of the lattice. 

\begin{figure}[htbp]
    \centering
    \includegraphics[scale = 1.05]{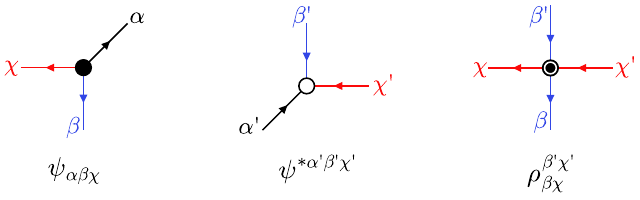}
    \caption{Graphical notation for the tripartite wavefunction $\psi$, its conjugate $\psi^*$ and the reduced density matrix $\rho_{BC}$. }
    \label{fig-wave-rho}
\end{figure}

\begin{figure}[htbp]
    \centering
    \includegraphics[scale = 1.05]{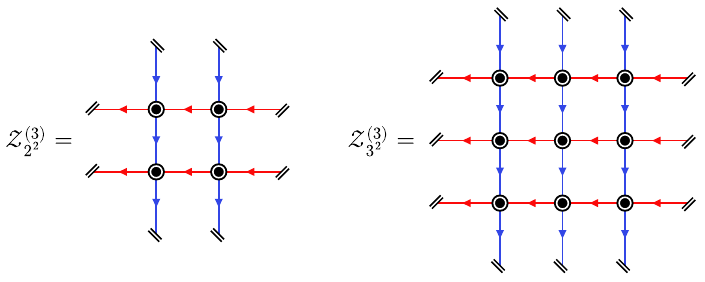}
    \caption{The construction of $\mathcal{Z}^{(\mathtt{q})}_{n^{\mathtt{q} - 1}}$ in the special case $\mathtt{q} = 3$ and $n = 2,3$. }
    \label{fig-replica-trick}
\end{figure}

\begin{figure}[htbp]
    \centering
    \includegraphics[scale = 1.05]{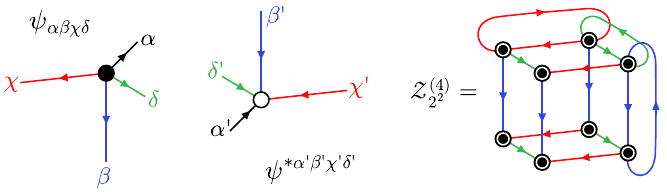}
    \caption{Graphical notation for the quadripartite wavefunction $\psi$, its conjugate $\psi^*$, and the construction of $\mathcal{Z}^{(\mathtt{q = 4})}_{2^{\mathtt{4} - 1}}$. Note that for clarity we omitted some identifications.}
    \label{fig-4-partite}
\end{figure}

For general $\mathtt{q}$-partite entanglement, the $n$-th R\'enyi multi-entropy is defined as\footnote{Compared with \cite{Gadde:2022cqi}, we use a slightly different notation here. We denote the partition function on the $n^{\mathtt{q - 1}}$-sheet Riemann surface by $\mathcal{Z}^{(\mathtt{q})}_{n^{\mathtt{q} - 1}}$ instead of $\mathcal{Z}^{(\mathtt{q})}_n$ in \cite{Gadde:2022cqi}. }
\begin{equation}\label{eq-RME-def}
S^{(\mathtt{q})}_n = \frac{1}{1 - n} \frac{1}{n^{\mathtt{q} - 2}} \log\Big(\mathcal{Z}^{(\mathtt{q})}_{n^{\mathtt{q} - 1}}/(\mathcal{Z}_{1}^{(\mathtt{q})})^{n^{\mathtt{q}-1}}\Big)\ ,
\end{equation}
where $\mathcal{Z}^{(\mathtt{q})}_{n^{\mathtt{q} - 1}}$ is the partition function obtained by contracting $n^{\mathtt{q} - 1}$ copies of the reduced density matrix with replica symmetry $\mathbb{Z}_n^{\otimes (\mathtt{q} - 1)}$ as described above. See Fig.~\ref{fig-wave-rho} for the graphical notation of the wavefunction and reduced density matrix, and Fig.~\ref{fig-replica-trick} for the contractions of reduced density matrices at $\mathtt{q} = 3$, $n = 2,3$. See also Fig.~\ref{fig-4-partite} for the generalization to quadripartite case with $n = 2$. Also note that when $\mathtt{q} = 2$, Eq.~\eqref{eq-RME-def} reduces to R\'enyi entropy. The multi-entropy is defined by taking the $n \to 1$ limit 
\begin{equation}\label{eq-multiE}
S^{(\mathtt{q})}= \lim_{n \to 1} S^{(\mathtt{q})}_{n}\ .
\end{equation}

For $\mathtt{q} = 3$ and $n=2$, the partition function contains 4 replicas with the gluing $\sigma_{B}=(13)(24)$ and $\sigma_{C}=(12)(34)$ as illustrated in Fig.~\ref{fig-replica-trick}. This results in a contraction of four density matrices (here we abbreviate $\rho_{BC}$ as $\rho$)
\begin{equation}\label{eq-replica}
\mathcal{Z}^{(3)}_{2^{3 - 1}} = \rho_{\beta_1 \chi_1}^{\beta_3 \chi_2} \rho_{\beta_2 \chi_2}^{\beta_4 \chi_1} \rho_{\beta_3 \chi_3}^{\beta_1 \chi_4} \rho_{\beta_4 \chi_4}^{\beta_2 \chi_3}\ .
\end{equation}
The corresponding R\'enyi tri-entropy\footnote{In this paper, the tripartite/quadripartite multi-entropy is also called ``tri/quadri-entropy''. } is given by
\begin{equation}\label{eq-ME2-def}
S^{(3)}_{2}(A;B;C) = \frac{1}{1 - 2}\frac{1}{2^{3-2}}\log\left(\mathcal{Z}^{(3)}_{2^{3 - 1}} / (\mathcal{Z}_{1}^{(3)})^{2^{3 - 1}}\right)\ ,
\end{equation}
with $\mathcal{Z}_{1}^{(3)} = \rho^{\beta\chi}_{\beta\chi}$ serving as the normalization factor.
\subsection{Genuine R\'enyi tri-entropy}\label{sec-GME-def}
The genuine R\'enyi multi-entropy $\mathrm{GM}_n^{(\mathtt{q})}$ is constructed based on the R\'enyi multi-entropy \eqref{eq-RME-def} and the R\'enyi entropy \eqref{eq-RenyiE}. It satisfies the following two properties~\cite{Iizuka:2025ioc,Iizuka:2025caq}: 
\begin{enumerate}[\textbullet]
\item $\mathrm{GM}_n^{(\mathtt{q})} (A_1; A_2; \cdots; A_{\mathtt{q}})$ is defined in terms of the $\mathtt{q}'$-partite R\'enyi multi-entropy $S_n^{(\mathtt{q}')}$ with $\mathtt{q}'\le \mathtt{q}$. 

\item $\mathrm{GM}_n^{(\mathtt{q})} (A_1; A_2; \cdots; A_{\mathtt{q}})$ vanishes for all states that can be factorized as
\begin{equation}\label{eq-GM-0}
\ket{\psi_{\mathtt{q}}}_{A_1\cdots A_{\mathtt{q}}} = \ket{\psi_{\tilde{\mathtt{q}}}}_{A'_1\cdots A'_{\tilde{\mathtt{q}}}}\otimes \ket{\psi_{\mathtt{q} - \tilde{\mathtt{q}}}}_{A'_{\tilde{\mathtt{q}} + 1}\cdots A'_{\mathtt{q}}}\ ,
\end{equation}
where $A'_i$ is a rearrangement of $A_i$, and $\tilde{\mathtt{q}} = 1, 2, \dots, \mathtt{q} - 1$. 
\end{enumerate}

When $\mathtt{q} = 3$, the genuine $n$-th R\'enyi tri-entropy is defined as~\cite{Iizuka:2025ioc,Iizuka:2025caq}
\begin{equation}\label{eq-GME}
\mathrm{GM}_n^{(3)}(A;B;C) = S_n^{(3)}(A;B;C) - \frac{1}{2}\big(S_n(A) + S_n(B) + S_n(C)\big)\ .
\end{equation}

There are two typical tripartite entangled states, GHZ-state and W-state:
\begin{equation}\label{eq-GHZ}
\ket{\text{GHZ}_k^N} = \sum\nolimits_{i = 0}^{k - 1} \ket{\underbrace{ii\cdots i}_{N}}/\sqrt{k}\ ,
\end{equation}
\begin{equation}\label{eq-W}
\ket{\text{W}^N} = \big(\ket{\underbrace{100\cdots 0}_N} + \ket{010\cdots 0} + \ket{001\cdots 0} + \cdots\big)/\sqrt{N}\ .
\end{equation}
We also consider the generalized GHZ-states and generalized W-states:
\begin{equation}\label{eq-gGHZ}
\ket{\widetilde{\text{GHZ}}_k^N} = \sum\nolimits_{i = 0}^{k - 1} \lambda_i \ket{ii\cdots i}\ ,
\end{equation}
\begin{equation}\label{eq-gW}
\ket{\widetilde{\text{W}}^N} = \cos\alpha \ket{100\cdots 0} + \sin\alpha \cos\beta \ket{010\cdots 0} + \sin\alpha \sin\beta \cos\chi \ket{001\cdots 0} + \cdots\ .
\end{equation}

When $n = 1$, the genuine tri-entropy is believed to capture the intrinsic tripartite entanglement.\footnote{In the case of CFTs, the genuine tri-entropy is associated with the OPE coefficients of the corresponding twist operators~\cite{Harper:2024ker}. } For the $\mathrm{GHZ}_k$-state \eqref{eq-GHZ}, 
\begin{equation}
\mathrm{GM}_1^{(3)}[\mathrm{GHZ}_k] = \frac{1}{2}\log k\ ,
\end{equation}
and for the generalized GHZ-states~\eqref{eq-gGHZ}, $\mathrm{GM}_1^{(3)} \ge 0$. Although we do not have a general proof that $\mathrm{GM}_1^{(3)}$ is always nonnegative, all the values of $\mathrm{GM}_1^{(3)}$ computed in this paper are indeed nonnegative.

When $n = 2$, the genuine R\'enyi tri-entropy can be interpreted as a symmetric version of the Markov gap. From the relation\footnote{We use the notation $AB$ to denote $A\cup B$.}
\begin{equation}\label{eq-relation}
S_2^{(3)}(A;B;C) = \frac{1}{2} S^R_{2,2}(A;B) + S_2(AB)
\end{equation}
between the second R\'enyi tri-entropy and the $(m,n)$-R\'enyi reflected entropy $S^R_{m,n}(A;B)$ of subsystems $A$ and $B$~\cite{Liu:2024ulq},\footnote{The relation~\eqref{eq-relation} can be verified using the replica trick for these quantities.} one finds that
\begin{equation}\label{eq-markov-similar}
\begin{split}
\mathrm{GM}_2^{(3)}(A;B;C) &= \frac{1}{2} S^R_{2,2}(A;B) + S_2(AB) - \frac{1}{2}\big(S_2(A) + S_2(B) + S_2(C)\big)\\
&= \frac{1}{2} \Big[S^R_{2,2}(A;B) - \big( S_2(A) + S_2(B) - S_2(AB) \big)\Big]\\
&= \frac{1}{2} \Big[S^R_{2,2}(A;B) - I_2(A;B)\Big]\ ,
\end{split}
\end{equation}
where $I_n(A;B) = S_n(A) + S_n(B) - S_n(AB)$ is the R\'enyi mutual information. The result in~\eqref{eq-markov-similar} closely resembles the definition of the Markov gap,
\begin{equation}
\Delta_n(A;B) = S_{1,n}^R(A;B) - I_n(A;B)\ ,
\end{equation}
evaluated at the R\'enyi index $n = 2$.

For the generalized W-state \eqref{eq-gW} with $N = 3$, the second genuine R\'enyi tri-entropy is nonnegative \cite{Harper:2025uui}, 
and for the generalized GHZ-state \eqref{eq-gGHZ},\footnote{See appendix~\ref{appx-GHZ} for a proof of \eqref{eq-GM2GHZ-le0}. }
\begin{equation}\label{eq-GM2GHZ-le0}
\mathrm{GM}_2^{(3)}[\widetilde{\text{GHZ}}_k^N]\le 0\ .
\end{equation}
\subsection{Genuine R\'enyi quadri-entropy}\label{sec-quadriE}
The genuine R\'enyi quadri-entropy is defined as \cite{Iizuka:2025ioc, Iizuka:2025caq}: 
\begin{equation}\label{eq-GM4}
\begin{split}
\mathrm{GM}_n^{(4)}&(A;B;C;D):=\, S_n^{(4)} (A;B;C;D)\\
&-\frac{1}{3}\Big(S_n^{(3)}(AB;C;D) + S_n^{(3)}(AC;B;D) + S_n^{(3)}(AD;B;C)\\
&\qquad\quad + S_n^{(3)}(BC;A;D) + S_n^{(3)}(BD;A;C) + S_n^{(3)}(CD;A;B)\Big)\\
& + a\Big( S_n(AB) + S_n(AC) + S_n(AD) \Big)\\
&+\Big(\frac{1}{3} - a\Big)\Big( S_n(A) + S_n(B) + S_n(C) + S_n(D) \Big)\ ,
\end{split}
\end{equation}
where $a$ is a real free parameter. For $\text{GHZ}_k$-state \eqref{eq-GHZ}, we have~\cite{Gadde:2022cqi}
\begin{equation}
S_n^{(\mathtt{q})}[\text{GHZ}_k] = \frac{n^{\mathtt{q} - 1} - 1}{(n - 1) n^{\mathtt{q} - 2}}\log k\ .
\end{equation}
Therefore, we have
\begin{equation}\label{eq-quadri-GHZ}
\begin{split}
\mathrm{GM}_n^{(4)}[\text{GHZ}_k] 
&= \frac{\log k}{n - 1}\Big[\frac{n^3 - 1}{n^2} - \frac{6}{3}\Big(\frac{n^2 - 1}{n}\Big) + 3a(n - 1) + 4\Big(\frac{1}{3} - a\Big)(n - 1)\Big]\\
&= \Big(\frac{3 - 3n + n^2}{3 n^2} - a\Big)\log k\ .
\end{split}
\end{equation}
Note that when $a = 1/12$, $\mathrm{GM}_2^{(4)}$ of the GHZ-state vanishes. One can also check that for the W-state \eqref{eq-W} with $N = 4$, we have
\begin{equation}\label{eq-quadri-W}
\mathrm{GM}_2^{(4)}[\text{W}^4] = a\log\frac{5^4}{2^9} + \log\frac{18}{5} - \frac{\log 5}{3} - \frac{\log 17}{4}\ ,
\end{equation}
which is positive when $a = 1/12$. Two other representative quadripartite entanglement patterns are the Cluster-state \cite{Briegel:2000unx} and the Dicke-state \cite{Stockton:2003nvk}, 
\begin{align}
\label{eq-cluster}
\ket{\text{Cluster}} &= \frac{1}{2}\big(\ket{0000} + \ket{0011} + \ket{1100} - \ket{1111}\big)\ ,\\
\label{eq-Dicke}
\ket{\text{Dicke}} &= \frac{1}{\sqrt{6}}\big(\ket{0011} + \ket{0101} + \ket{0110} + \ket{1001} + \ket{1010} + \ket{1100}\big)\ .
\end{align}
The $n = 2$ genuine R\'enyi quadri-entropy for the Cluster-state~\eqref{eq-cluster} is 
\begin{equation}\label{eq-Cluster-GM2}
\mathrm{GM}_2^{(4)}[\text{Cluster}] = \Big(a - \frac{1}{12}\Big) \log 2\ ,
\end{equation}
which vanishes when $a = 1/12$. For the Dicke-state~\eqref{eq-Dicke},  
\begin{equation}\label{eq-Dicke-GM2}
\mathrm{GM}_2^{(4)}[\text{Dicke}] = \Big(\frac{1}{12} - a\Big) \log 2 - \frac{\log(3^7\times 5\times 11)}{4} + \log 19\ ,
\end{equation}
which is positive when $a = 1/12$. 
\section{Genuine multi-entropy for torus link states}\label{sec-GME-links}
\subsection{Multi-boundary states in Chern-Simons theory}
We consider Chern-Simons theory with gauge group $G$ at level $k$, whose action on a 3-manifold $\mathcal{M}$ is
\begin{equation}\label{eq-CS-action}
S_{\text{CS}}[A] = \frac{k}{4\pi}\int_\mathcal{M}\mathrm{Tr}\Big(A\wedge \mathrm{d}A + \frac{2}{3}A\wedge A\wedge A\Big)\ ,
\end{equation}
with $A$ the gauge field. In this paper we study a class of states called \textit{link states}. These states are defined on $\Sigma_N$, which is the disjoint union of $N$ copies of $\mathbb{T}^2$, 
\begin{equation}
\Sigma_N := \bigcup\nolimits_{i = 1}^N \mathbb{T}^2\ .
\end{equation}

\begin{figure}[htbp]
    \centering
    \includegraphics[scale = 1.2]{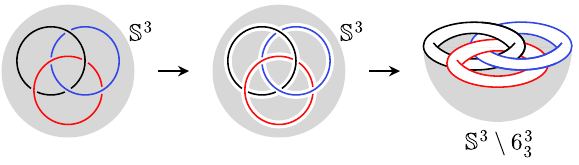}
    \caption{ Figure showing the construction of link $6_3^3$ complement. Removing a tubular neighborhood around link $6_3^3$ embedded in $\mathbb{S}^3$ results in link $6_3^3$ complement which is a manifold with three torus boundaries. We note that the color in this figure (and Fig.~\ref{fig-three-links}) is used to distinguish different components and does not denote specific representation. }
    \label{fig-setup}
\end{figure}

One can prepare these states by performing the Euclidean path integral of the theory~\eqref{eq-CS-action} on a 3-manifold $\mathcal{M}$ whose boundary is $\Sigma_N$. In this paper, we focus on the following construction. We start from $\mathbb{S}^3$ containing a generic $N$-component link $\mathcal{L}^N$, which consists of $N$ circles, i.e., $\mathcal{L}^N = \mathcal{C}_1 \cup \mathcal{C}_2 \cup \cdots \cup \mathcal{C}_N$ (see the left part of Fig.~\ref{fig-setup}). We then remove a tubular neighborhood of $\mathcal{L}^N$ (middle part of Fig.~\ref{fig-setup}), obtaining a 3-manifold denoted by $\mathcal{M}_N = \mathbb{S}^3 \backslash \mathcal{L}^N$, whose boundary is $\Sigma_N$. The construction for the case $N=3$ with $\mathcal{L}^3 = 6_3^3$ is illustrated in Fig.~\ref{fig-setup}.

The wavefunction of the link state constructed in the above manner is given by~\cite{Balasubramanian:2016sro}
\begin{equation}
\ket{\mathcal{L}^N} = \sum_{j_{1,2,\dots,N}} C_{\mathcal{L}^N}(j_1,j_2,\dots,j_N)\ket{j_1,j_2,\dots,j_N}\ ,
\end{equation}
where the complex coefficients $C_{\mathcal{L}^N}(j_1,j_2,\dots,j_N)$ are the colored link invariants of Chern-Simons theory, with the representation $R_{j_i}^*$ assigned to the $i$-th component of the link,
\begin{equation}\label{eq-wave-coefficient}
\begin{split}
C_{\mathcal{L}^N}(j_1,j_2,\dots,j_N) &= \braket{j_1,j_2,\dots,j_N|\mathcal{L}^N}\\
&= \left< W_{R^*_{j_1}}\!(\mathcal{C}_1)\, W_{R^*_{j_2}}\!(\mathcal{C}_2)\cdots W_{R^*_{j_N}}\!(\mathcal{C}_N)\right>\ .
\end{split}
\end{equation}

\subsection{Abelian case}
In this section, we focus on link states in $\mathrm{U}(1)_k$ Chern-Simons theory and study their genuine R\'enyi multi-entropies. For the three-component link case (Section~\ref{subsec-abelian-3Link}), we obtain an analytical expression~\eqref{eq-n-GME-3link} valid for a general R\'enyi index $n$. For $N$-component links with $N > 3$, analytical results are available only for $n = 2$.

\subsubsection{Three-component links}\label{subsec-abelian-3Link}
For a generic three-component link $\mathcal{L}^3$ in $\mathrm{U}(1)_k$ Chern-Simons theory, the corresponding normalized wavefunction can be expressed in terms of the colored link invariants~\cite{Witten:1988hf}\footnote{The $\mathrm{U}(1)_k$ Chern-Simons link states \eqref{eq-wave3} and \eqref{eq-waveN} are also called \textit{weighted graph states}. See \cite{Hein:2006uvf, hein2004multiparty, schlingemann2003cluster} for graph states on qubits and \cite{tang2013greenberger, chen2016graph} for $k$-bits. }
\begin{equation}\label{eq-wave3}
\ket{\mathcal{L}^3} = \frac{1}{k^{3/2}}\sum_{\alpha,\beta,\chi = 0}^{k - 1} e^{\frac{2\pi i}{k}(\alpha \beta L_{AB} + \beta \chi L_{BC} + \chi \alpha L_{CA})}\ket{\alpha,\beta,\chi}\ ,
\end{equation}
where $\alpha$, $\beta$, and $\chi$ denote the charges assigned to the loops $A$, $B$, and $C$, respectively, and $L_{AB}$ is the Gauss linking number between loops $A$ and $B$ (satisfying $L_{AB} = L_{BA}$); similarly for $L_{BC}$ and $L_{CA}$.
The reduced density matrix $\rho_{BC}$ is obtained by tracing out subsystem $A$, and its matrix elements are given by
\begin{equation}\label{eq-rhoBC}
\begin{split}
\bra{\beta,\chi}\rho_{BC}\ket{\beta',\chi'} &= \sum_{\alpha = 0}^{k - 1} \braket{\alpha,\beta,\chi|\mathcal{L}^3}\braket{\mathcal{L}^3|\alpha,\beta',\chi'}\\
&=\frac{1}{k^3} e^{\frac{2\pi i}{k}(\beta \chi - \beta' \chi')L_{BC}}\sum_{\alpha}e^{\frac{2\pi i}{k}\alpha((\beta - \beta')L_{AB} + (\chi - \chi')L_{CA})}\\
&=\frac{1}{k^2} e^{\frac{2\pi i}{k}(\beta \chi - \beta' \chi')L_{BC}} \eta\!\left[(\beta-\beta')L_{AB}+(\chi-\chi')L_{CA}\right]\!\ ,
\end{split}
\end{equation}
where we have introduced the $\eta$-constraint function,\footnote{The notation $x \equiv y\ (\mathrm{mod}\ k)$ denotes that the two integers $x$ and $y$ are congruent modulo $k$.}
\begin{equation}\label{eq-def-eta}
\eta[x] = \frac{1}{k} \sum_{\alpha = 0}^{k - 1} e^{\frac{2\pi i}{k}\alpha x} = 
\begin{cases}
1\ , & \text{if}\ \ x \equiv 0\ (\mathrm{mod}\ k)\ ,\\
0\ , & \text{otherwise.}
\end{cases}
\end{equation}
The $\eta$-constraint function~\eqref{eq-def-eta} satisfies the following properties:
\begin{equation}\label{eq-property}
\eta[x] = \eta[-x]\ , \quad \eta[x]^2 = \eta[x]\ , \quad \eta[x]\,\eta[y] = \eta[x + y]\,\eta[x]\ .
\end{equation}

With the reduced density matrix~\eqref{eq-rhoBC}, the second R\'enyi multi-entropy among $A$, $B$, and $C$ can be computed from its definitions~\eqref{eq-replica} and~\eqref{eq-ME2-def}. We first evaluate the partition function on the replica space using~\eqref{eq-replica},
\begin{equation}\label{eq-3linkZ}
\begin{split}
\mathcal{Z}_4^{(3)} = \frac{1}{k^{8}}&\sum_{\beta_{1\sim 4},\chi_{1\sim 4}}\!\Big\{\!\exp\!\Big[\frac{2\pi i}{k}\big((\beta_1 - \beta_4)(\chi_1 - \chi_4) + (\beta_2 - \beta_3)(\chi_2 - \chi_3)\big)L_{BC}\Big] \\
&\times \eta\big[(\beta_1-\beta_3)L_{AB}+(\chi_1-\chi_2)L_{CA}\big]\ \eta\big[(\beta_2-\beta_4)L_{AB}+(\chi_2-\chi_1)L_{CA}\big]\\
&\times \eta\big[(\beta_3-\beta_1)L_{AB}+(\chi_3-\chi_4)L_{CA}\big]\ \eta\big[(\beta_4-\beta_2)L_{AB}+(\chi_4-\chi_3)L_{CA}\big]\!\Big\}\ ,
\end{split}
\end{equation}
which is a sum of terms of the form $e^{(\cdots)}/k^8$ over $\beta_{1,2,3,4}$ and $\chi_{1,2,3,4}$, subject to the constraints imposed by the $\eta$-functions. To handle these $\eta$-constraints, note that, for example, $\eta\big[(\beta_1-\beta_3)L_{AB}+(\chi_1-\chi_2)L_{CA}\big]$ is a function of\footnote{We use the notation ``$:=$'' to denote a definition. }
\begin{equation}
\beta_{13}:=(\beta_1 - \beta_3)\ \mathrm{mod}\ k\quad \text{and}\quad \chi_{12}:= (\chi_1 - \chi_2)\ \mathrm{mod}\ k\ . 
\end{equation}
Therefore, we can make the following change of variables
\begin{equation}
\beta_3 \to \beta_1 - \beta_{13}\ , \quad \beta_4 \to \beta_2 - \beta_{24}\ ,\quad \chi_2 \to \chi_1 - \chi_{12}\ , \quad \chi_4 \to \chi_3 - \chi_{34}
\end{equation}
in \eqref{eq-3linkZ} and obtain
\begin{equation}\label{eq-Z-1}
\begin{split}
\mathcal{Z}_4^{(3)} = \frac{1}{k^{8}}&\sum_{\beta_{1,2,13,24},\chi_{1,3,12,34}}\!\Big\{\!\exp\!\Big[\frac{2\pi i}{k}(-\beta_{13}\chi_{12} + \beta_{24}\chi_{34})L_{BC}\Big] \\
&\times \exp\!\Big[\frac{2\pi i}{k}\beta_1(\chi_{12} + \chi_{34})L_{BC}\Big] \exp\!\Big[\frac{2\pi i}{k}\beta_2 (-\chi_{12} - \chi_{34})L_{BC}\Big]\\
&\times \exp\!\Big[\frac{2\pi i}{k}\chi_1 (\beta_{13} + \beta_{24})L_{BC}\Big] \exp\!\Big[\frac{2\pi i}{k}\chi_3 (-\beta_{13} - \beta_{24})L_{BC}\Big]\\
&\times \eta\big[\beta_{13}L_{AB}+\chi_{12}L_{CA}\big]\, \eta\big[\beta_{24}L_{AB}-\chi_{12}L_{CA}\big]\\
&\times \eta\big[-\beta_{13}L_{AB}+\chi_{34}L_{CA}\big]\, \eta\big[-\beta_{24}L_{AB}-\chi_{34}L_{CA}\big]\!\Big\}\ .
\end{split}
\end{equation}
Since $\beta_{1,2}$ and $\chi_{1,3}$ do not appear in the $\eta$-constraints in \eqref{eq-Z-1}, we can sum them using \eqref{eq-def-eta} and end up with a summation over $\beta_{13,24}, \chi_{12,34}$ of 
\begin{equation}
\frac{1}{k^{4}} \exp\!\left[\frac{2\pi i}{k} (-\beta_{13}\chi_{12} + \beta_{24}\chi_{34})L_{BC}\right]
\end{equation}
with the following constraints
\begin{equation}\label{eq-constraint}
\begin{split}
1=&\,\eta\big[(\chi_{12} + \chi_{34})L_{BC}\big]\, \eta\big[(-\chi_{12} - \chi_{34})L_{BC}\big]\, \eta\big[(\beta_{13} + \beta_{24})L_{BC}\big]\, \eta\big[(-\beta_{13} - \beta_{24})L_{BC}\big]\\
&\times \eta\big[\beta_{13}L_{AB}+\chi_{12}L_{CA}\big]\, \eta\big[\beta_{24}L_{AB}-\chi_{12}L_{CA}\big]\\
&\times\eta\big[-\beta_{13}L_{AB}+\chi_{34}L_{CA}\big]\, \eta\big[-\beta_{24}L_{AB}-\chi_{34}L_{CA}\big]\ .
\end{split}
\end{equation}
Among them, the constraints $1 = \eta\big[(\chi_{12} + \chi_{34})L_{BC}\big]$ and $1 = \eta\big[(\beta_{13} + \beta_{24})L_{BC}\big]$ force
\begin{equation}
\exp\!\left[\frac{2\pi i}{k} (-\beta_{13}\chi_{12} + \beta_{24}\chi_{34})L_{BC}\right] = 1\ .
\end{equation}
Therefore, the summation reduces to finding out how many sets of $(\beta_{13,24},\chi_{12,34})\in \mathbb{Z}_k^{\otimes 4}$ can survive under the constraint \eqref{eq-constraint}. To figure this out, we first use the properties in \eqref{eq-property} to simplify \eqref{eq-constraint} to
\begin{equation}\label{eq-constraint-simplified}
\begin{split}
1=&\,\eta\big[(\chi_{12} + \chi_{34})L_{BC}\big]\ \eta\big[(\beta_{13} + \beta_{24})L_{BC}\big]\\
&\times \eta\big[(\beta_{13} + \beta_{24})L_{AB}\big]\ \eta\big[(\chi_{12} + \chi_{34})L_{CA}\big]\ \eta\big[\beta_{13}L_{AB}+\chi_{12}L_{CA}\big]\ .
\end{split}
\end{equation}
Performing the change of variables
$$
\beta_{24} \to b - \beta_{13}\ ,\quad \chi_{34} \to c - \chi_{12}\ ,
$$
\eqref{eq-constraint-simplified} is further simplified to
\begin{equation}\label{eq-constraint-full-simplified}
\begin{split}
1 = \eta\big[c\,L_{BC}\big]\, \eta\big[c\,L_{CA}\big]\, \eta\big[b\,L_{BC}\big]\, \eta\big[b\,L_{AB}\big]\, \eta\big[\beta_{13}L_{AB}+\chi_{12}L_{CA}\big]\ .
\end{split}
\end{equation}
Now we can count the number of solutions. There are (``$\gcd$'' denotes the greatest common divisor)
\begin{enumerate}[\textbullet]
\item $\gcd(k,L_{AB},L_{BC})$ possible $b\in \mathbb{Z}_k$ satisfying $1 = \eta\big[b\,L_{BC}\big]\, \eta\big[b\,L_{AB}\big]$, 
\item $\gcd(k,L_{BC},L_{CA})$ possible $c\in \mathbb{Z}_k$ satisfying $1= \eta\big[c\,L_{BC}\big]\,\eta\big[c\,L_{CA}\big]$,
\item $k\cdot\gcd(k,L_{CA},L_{AB})$ possible $\{\beta_{13}, \chi_{12}\}\in \mathbb{Z}_k^{\otimes 2}$ satisfying $1= \eta\big[\beta_{13}L_{AB}+\chi_{12}L_{CA}\big]$.\footnote{See footnote 3 of \cite{Balasubramanian:2016sro}.}
\end{enumerate}
Therefore, we obtain
\begin{equation}\label{eq-3-ME2}
\begin{split}
&\mathcal{Z}_4^{(3)} = k^{-3} \gcd(k,L_{CA},L_{AB}) \gcd(k,L_{AB},L_{BC}) \gcd(k,L_{BC},L_{CA})\\
\Rightarrow\, & S_{2}^{(3)}(A;B;C) = \frac{1}{2}\log\frac{k^3}{\gcd(k,L_{CA},L_{AB}) \gcd(k,L_{AB},L_{BC}) \gcd(k,L_{BC},L_{CA})}\ .
\end{split}
\end{equation}
One can see that the key idea is to use the variable substitution together with~\eqref{eq-def-eta} to generate as many constraints as possible, so that the summand becomes $1$, thereby reducing the problem to counting the number of solutions that satisfy all the constraints. However, this task becomes difficult when the R\'enyi index $n>2$. Nevertheless, if one of the three linking numbers is set to zero, the problem can be solved (without loss of generality, we set $L_{BC} = 0$ here),
\begin{align}
\label{eq-improve}
S_n^{(3)}(A;B;C)|_{L_{BC} = 0} =&\, \frac{1}{n}\log\frac{k^{n + 1}}{\gcd(k,L_{CA},L_{AB})^{n - 1} \gcd(k,L_{AB}) \gcd(k,L_{CA})}\\
\label{eq-improve1} 
=&\, \frac{1}{n}\log\frac{k^{3}}{\gcd(k,L_{CA},L_{AB}) \gcd(k,L_{AB}) \gcd(k,L_{CA})}\\
\label{eq-improve2}
&\, + \Big(1 - \frac{2}{n}\Big) \log\frac{k}{\gcd(k,L_{CA},L_{AB})}\ .
\end{align}
We leave the tedious derivation of~\eqref{eq-improve} to Appendix~\ref{appx-derive}. We now aim to obtain a complete result for $L_{BC} \neq 0$, based on the $n = 2$ result~\eqref{eq-3-ME2} and the general-$n$, $L_{BC} = 0$ result~\eqref{eq-improve}. By comparing~\eqref{eq-improve1} with~\eqref{eq-3-ME2} and recalling that $L_{BC} = 0$, we are led to perform the following improvements in \eqref{eq-improve1}:
\begin{equation}
\begin{split}
&\gcd(k,L_{AB}) = \gcd(k, L_{AB}, 0) \to \gcd(k,L_{AB}, L_{BC})\ ,\\
&\gcd(k,L_{CA}) = \gcd(k, 0, L_{CA}) \to \gcd(k,L_{BC}, L_{CA})\ .
\end{split}
\end{equation}
By symmetry among $L_{AB}$, $L_{BC}$, and $L_{CA}$, the term $\gcd(k,L_{CA},L_{AB})$ in~\eqref{eq-improve2} should be replaced with $\gcd(k,L_{CA},L_{AB},L_{BC})$. This leads to a formula for the $n$-th R\'enyi multi-entropy of three-component links with linking numbers ${L_{AB}, L_{BC}, L_{CA}}$
\begin{equation}\label{eq-n-multiE-3link}
\begin{split}
S_n^{(3)} (A;B;C) =&\, \frac{1}{n}\log\frac{k^{3}}{\gcd(k,L_{CA},L_{AB}) \gcd(k,L_{AB},L_{BC}) \gcd(k,L_{BC},L_{CA})}\\
&\, + \Big(1 - \frac{2}{n}\Big) \log\frac{k}{\gcd(k,L_{CA},L_{AB},L_{BC})}\ .
\end{split}
\end{equation}
We emphasize that the formula~\eqref{eq-n-multiE-3link} above is a conjecture, based on the rigorous results~\eqref{eq-3-ME2} and~\eqref{eq-improve}. Nevertheless, it has been confirmed by extensive numerical checks.\footnote{We numerically verified~\eqref{eq-n-multiE-3link} for $n = 3$ for all three-component links with $k \le 17$, for $n = 4$ with $k \le 6$, for $n = 5$ with $k \le 5$, and for $n = 6$ with $k \le 4$.}

The $n$-th R\'enyi entropy $S_n(A;BC)$ for the wavefunction~\eqref{eq-wave3} is given by~\cite{Balasubramanian:2016sro}
\begin{equation}\label{eq-3-RE2}
S_n(A;BC) = \log\frac{k}{\gcd(k,L_{CA},L_{AB})}\ ,
\end{equation}
from which it follows that the genuine R\'enyi multi-entropy~\eqref{eq-GME} of the three-component link state \eqref{eq-wave3} is
\begin{equation}\label{eq-n-GME-3link}
\begin{split}
\mathrm{GM}_n^{(3)} (A;B;C) = \Big(\frac{1}{n} - \frac{1}{2}\Big)\log\frac{k\cdot \gcd(k,L_{CA},L_{AB},L_{BC})^2}{\gcd(k,L_{CA},L_{AB}) ,\gcd(k,L_{AB},L_{BC}) ,\gcd(k,L_{BC},L_{CA})}\ .
\end{split}
\end{equation}
In Appendix~\ref{appx-Veen}, we prove the following inequality using a Venn diagram:
\begin{equation}\label{eq-ineq}
1 \le \frac{k\cdot \gcd(k,L_{CA},L_{AB},L_{BC})^2}{\gcd(k,L_{CA},L_{AB}) ,\gcd(k,L_{AB},L_{BC}) ,\gcd(k,L_{BC},L_{CA})} \le k\ ,
\end{equation}
from which it is clear that the logarithmic term in~\eqref{eq-n-GME-3link} is nonnegative and bounded above by $\log k$.
\subsubsection{\boldmath $N$-component links\unboldmath}\label{sec-Nlink}
The calculation in Section~\ref{subsec-abelian-3Link} can be directly generalized to $N$-component links\footnote{Here, $N$ denotes the number of components, while $n$ always refers to the R\'enyi index.} and we present only an abbreviated derivation. We divide the $N$ loops into three groups, $\{A_1, A_2, \dots, A_{N_A}\}$, $\{B_1, B_2, \dots, B_{N_B}\}$ and $\{C_1, C_2, \dots, C_{N_C}\}$, with $L_{A_i B_j}$, $L_{B_i C_j}$ and $L_{C_i A_j}$ the corresponding Gauss linking numbers between different groups\footnote{There can be non-trivial linking numbers between loops in the same group like $L_{A_i A_j}$. However, both the R\'enyi multi-entropy and the R\'enyi entropy do not depend on them~\cite{Balasubramanian:2018por}. Therefore, we do not explicitly include terms such as $\alpha_i \alpha_j L_{A_i A_j}$ in \eqref{eq-waveN}. }. The wavefunction is
\begin{equation}\label{eq-waveN}
\ket{\mathcal{L}^N} = \frac{1}{k^{N/2}}\!\!\!\sum_{\alpha_i,\beta_j,\chi_k = 0}^{k - 1}\!\!\! e^{\frac{2\pi i}{k}\sum_{i,j}(\alpha_i \beta_j L_{A_i B_j} + \beta_i \chi_j L_{B_i C_j} + \chi_i \alpha_j L_{C_i A_j} +\, \cdots)}\ket{\{\alpha_i\}, \{\beta_j\}, \{\chi_k\}}\ .
\end{equation}
The second R\'enyi multi-entropy can be computed similarly:
\begin{equation}\label{eq-N-com-ME2}
S_2^{(3)}(A;B;C) = \frac{1}{2}\log\frac{k^N}{|\ker G_{BC,A}|\cdot |\ker G_{CA,B}|\cdot|\ker G_{AB,C}|}\ ,
\end{equation}
where $G_{BC,A}$ is a $(N_B + N_C)\times N_A$-matrix called linking matrix
\begin{equation}
G_{BC,A} = 
\begin{pmatrix}
L_{B_1 A_1} & L_{B_1 A_2} & \cdots & L_{B_1 A_{N_A}}\\
\vdots & \vdots & \ddots & \vdots \\
L_{B_{N_B} A_1} & L_{B_{N_B} A_2} & \cdots & L_{B_{N_B} A_{N_A}}\\
L_{C_1 A_1} & L_{C_1 A_2} & \cdots & L_{C_1 A_{N_A}}\\
\vdots & \vdots & \ddots & \vdots \\
L_{C_{N_C} A_1} & L_{C_{N_C} A_2} & \cdots & L_{C_{N_C} A_{N_A}}
\end{pmatrix}
\end{equation}
and $|\ker G_{BC,A}|$ denotes the number of solutions $\Vec{x}\in \mathbb{Z}_k^{\otimes N_A}$ to the
congruence equations 
\begin{equation}
G_{BC,A}\cdot \Vec{x} \equiv 0\ (\mmod k)\ .    
\end{equation}

The $n$-th R\'enyi entropy $S_n(A;BC)$ for this general $N$-component link is given by~\cite{Balasubramanian:2016sro}, 
\begin{equation}\label{eq-N-RE}
S_n(A;BC) = \log\frac{k^{N_A}}{|\ker G_{BC,A}|}\ .
\end{equation}
One can see that for $N$-links, $\mathrm{GM}_2^{(3)}$ vanishes just as in the three-component case~\eqref{eq-n-GME-3link}.\footnote{In this paper we sometimes abbreviate ``$N$-component link'' as ``$N$-link''.}
\subsubsection{Quadri-entropy of four-component links}
We also compute $S_2^{(4)}(A;B;C;D)$ using the method described in Section~\ref{subsec-abelian-3Link}:
\begin{equation}\label{eq-ME42}
\begin{split}
S_2^{(4)}(A;B;C;D) =&\, \frac{1}{4}\log\frac{k^4}{|\ker G_{BCD,A}|\cdot|\ker G_{CDA,B}|\cdot|\ker G_{DAB,C}|\cdot|\ker G_{ABC,D}|}\\
&+\frac{1}{4}\log\frac{k^6}{|\ker G_{AB,CD}|\cdot|\ker G_{AC,BD}|\cdot|\ker G_{AD,BC}|}\\
=&\, \frac{1}{4}\Big(S_2(A) + S_2(B) + S_2(C) + S_2(D) + S_2(AB) + S_2(AC) + S_2(AD)\Big)\ .
\end{split}
\end{equation}
One can check that if $a = 1/12$, the corresponding second genuine R\'enyi quadri-entropy $\mathrm{GM}_2^{(4)}$~\eqref{eq-GM4} vanishes, similar to the second genuine R\'enyi  tri-entropy $\mathrm{GM}_2^{(3)}$ case~\eqref{eq-n-GME-3link}, implying that four-link states contain only GHZ-type and Cluster-type quadripartite entanglement (see the results \eqref{eq-quadri-GHZ}, \eqref{eq-quadri-W}, \eqref{eq-Cluster-GM2} and \eqref{eq-Dicke-GM2} in Section~\ref{sec-quadriE}). 
\subsection{Non-Abelian case}
\begin{figure}[htbp]
    \centering
    \includegraphics[scale = 1]{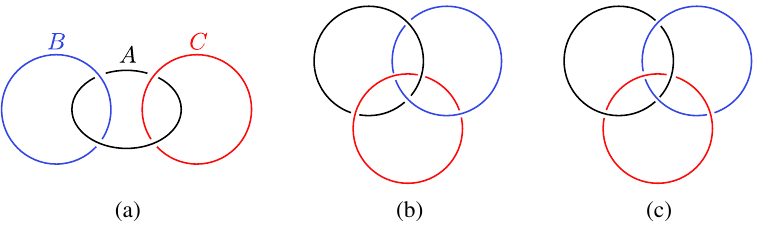}
    \caption{\label{fig-three-links}(a) The connected sum of two Hopf links (the link $2_1^2 + 2_1^2$). (b) The link $6_3^3$. (c) Borromean rings (the link $6_2^3$). }
\end{figure}

In this section, we focus on $\mathrm{SU}(2)_k$ Chern-Simons theory. We consider three-component links (see Fig.~\ref{fig-three-links}) and compute the genuine tri-entropy among the components. 
\subsubsection{\boldmath Connected sum of two Hopf links\unboldmath}
We start with the link $2_1^2 + 2_1^2$, which is the connected sum of two Hopf links (see Fig.~\ref{fig-three-links}~(a)). By evaluating the link invariant in Eq.~\eqref{eq-wave-coefficient}, one obtains the wavefunction of this link~\cite{Balasubramanian:2016sro}
\begin{equation}\label{eq-hopf-hopf}
\ket{2_1^2 + 2_1^2} = \sum_{\alpha,\beta,\chi,m} S_{\alpha m}N_{m \beta \chi}\ket{\alpha,\beta,\chi} = \sum_{\alpha,\beta,\chi}\frac{S_{\beta\alpha} S_{\chi\alpha}}{S_{0 \alpha}}\ket{\alpha,\beta,\chi}\ ,
\end{equation}
where the spins $\alpha = 0, \tfrac{1}{2}, \dots, \tfrac{k}{2}$ label the integrable representations of $\mathrm{SU}(2)_k$, $N_{m \beta \chi}$ denotes the fusion coefficients, and at the second equal sign we have applied the Verlinde formula
\begin{equation}
N_{m \beta \chi} = \sum_j \frac{S_{m j}S_{\beta j}S_{\chi j}}{S_{0 j}}
\end{equation}
and the property 
\begin{equation}\label{eq-property-S}
S_{\alpha m}S_{m j} = S^*_{\bar{\alpha}m}S_{m j} = \delta_{\bar{\alpha} j}
\end{equation}
of the $S$-matrix\footnote{The specific form of the $\mathrm{SU}(2)_k$ $S$-matrix is
\begin{equation}
S_{\alpha\beta} = \sqrt{\frac{2}{k + 2}} \sin\!\left(\!\frac{(2\alpha + 1)(2\beta + 1)\pi}{k + 2}\!\right)\!\ .
\end{equation}
}. The reduced density matrix is given by\footnote{Note that the reduced density matrix $\rho_{BC}$ in \eqref{eq-hopf-rhoBC} is unnormalized, $\mathrm{Tr}\rho_{BC} = \sum_\alpha |S_{0 \alpha}|^{-2} = \sum_\alpha \mathcal{D}^2/d_\alpha^2$. }
\begin{equation}\label{eq-hopf-rhoBC}
\bra{\beta,\chi}\rho_{BC}\ket{\beta',\chi'} = \sum_\alpha \frac{S_{\beta\alpha} S_{\chi\alpha}}{S_{0\alpha}} \frac{S^*_{\beta'\alpha} S^*_{\chi'\alpha}}{S^*_{0\alpha}}\ .
\end{equation}
Using \eqref{eq-replica}, we get (repeated indices in the following expression are to be summed over)
\begin{equation}
\begin{split}
\mathcal{Z}_4^{(3)} =&\, \frac{{\color{blue}S_{\beta_1\alpha_1}} {\color{red}S_{\chi_1\alpha_1}}}{S_{0\alpha_1}} \frac{S^*_{\beta_3\alpha_1} S^*_{\chi_2\alpha_1}}{S^*_{0\alpha_1}}
\frac{S_{\beta_2\alpha_2} S_{\chi_2\alpha_2}}{S_{0\alpha_2}} \frac{S^*_{\beta_4\alpha_2} {\color{red}S^*_{\chi_1\alpha_2}}}{S^*_{0\alpha_2}}\\
&\times
\frac{S_{\beta_3\alpha_3} S_{\chi_3\alpha_3}}{S_{0\alpha_3}} \frac{{\color{blue}S^*_{\beta_1\alpha_3}} S^*_{\chi_4\alpha_3}}{S^*_{0\alpha_3}}
\frac{S_{\beta_4\alpha_4} S_{\chi_4\alpha_4}}{S_{0\alpha_4}} \frac{S^*_{\beta_2\alpha_4} S^*_{\chi_3\alpha_4}}{S^*_{0\alpha_4}}
\end{split}
\end{equation}
Using the property \eqref{eq-property-S} of $S$-matrix repeatedly, one can sum over $\beta_{1,2,3,4}$, $\chi_{1,2,3,4}$ and end up with $\delta$-functions
\begin{equation}\label{eq-212-212-entropy}
\begin{split}
\mathcal{Z}_4^{(3)} &= \frac{{\color{blue}\delta_{\alpha_1\alpha_3}}{\color{red}\delta_{\alpha_1\alpha_2}}}{S_{0\alpha_1}} \frac{\delta_{\alpha_1\alpha_3} \delta_{\alpha_1\alpha_2}}{S^*_{0\alpha_1}}
\frac{\delta_{\alpha_2\alpha_4}}{S_{0\alpha_2}} \frac{\delta_{\alpha_2\alpha_4}}{S^*_{0\alpha_2}}\frac{\delta_{\alpha_3\alpha_4}}{S_{0\alpha_3}} \frac{\delta_{\alpha_3\alpha_4}}{S^*_{0\alpha_3}}
\frac{1}{S_{0\alpha_4}} \frac{1}{S^*_{0\alpha_4}}\\
&=\sum_{\alpha} |S_{0\alpha}|^{-8} = \sum_\alpha \mathcal{D}^8\!/d_\alpha^8\\
\Rightarrow S_2^{(3)}(A;B;C) &= -\frac{1}{2}\log \frac{\sum_\alpha \mathcal{D}^8\!/d_\alpha^8}{\left(\sum_\alpha \mathcal{D}^2\!/d_\alpha^2\right)^4} = \frac{1}{2}\log\frac{\left(\sum_\alpha d_\alpha^{-2}\right)^4}{\sum_\alpha d_\alpha^{-8}}\ ,
\end{split}
\end{equation}
where in the second line we sum over $\alpha_{2,3,4}$ and rename $\alpha_1$ as $\alpha$. In the final result, $d_\alpha$ is the quantum dimension of anyon $\alpha$ and 
\begin{equation}
\mathcal{D} = \sqrt{\sum\nolimits_\alpha d_\alpha^2}
\end{equation}
is the total quantum dimension. 

One can calculate the $n$-th R\'enyi multi-entropy similarly, and the result is
\begin{equation}\label{eq-212-212-n-entropy}
\begin{split}
\mathcal{Z}_{n^2}^{(3)} &= \sum\nolimits_{\alpha} |S_{0 \alpha}|^{-2 n^2} = \sum\nolimits_\alpha \mathcal{D}^{2 n^2}\!/d_\alpha^{2 n^2}\\
\Rightarrow S_n^{(3)}(A;B;C) &= \frac{1}{1-n}\frac{1}{n}\log \frac{\sum_\alpha \mathcal{D}^{2 n^2}\!/d_\alpha^{2 n^2}}{\left(\sum_\alpha \mathcal{D}^2\!/d_\alpha^2\right)^{n^2}} = \frac{\log\big(\sum\nolimits_\alpha p_\alpha^{n^2}\big)}{(1 - n)n}\ , \quad p_\alpha = \frac{d_\alpha^{-2}}{\sum_j d_j^{-2}}\ .
\end{split}
\end{equation}
In the $n\to 1$ limit, we have
\begin{equation}\label{eq-212-212-1-entropy}
\begin{split}
S_1^{(3)}(A;B;C) &= \lim_{n\to 1}\frac{1}{n - 1}\Bigg[\log\frac{\left(\sum_\alpha d_\alpha^{-2}\right)^{n^2}}{\sum_\alpha d_\alpha^{-2 n^2}} - \log\frac{\sum_\alpha d_\alpha^{-2}}{\sum_\alpha d_\alpha^{-2}} + \log\frac{\sum_\alpha d_\alpha^{-2}}{\sum_\alpha d_\alpha^{-2}}\Bigg]\\
&=\lim_{n\to 1}\frac{\partial}{\partial n} \log\frac{\left(\sum_\alpha d_\alpha^{-2}\right)^{n^2}}{\sum_\alpha d_\alpha^{-2 n^2}}\\
&= - 2 \sum\nolimits_\alpha p_\alpha \log p_\alpha\ .
\end{split}
\end{equation}

The $n$-th R\'enyi entropy of the state~\eqref{eq-hopf-hopf} is~\cite{Balasubramanian:2016sro}
\begin{equation}\label{eq-633-RE2}
S_n(AB;C) = S_n(BC;A) = S_n(CA;B) = \frac{\log\big(\sum\nolimits_\alpha p_\alpha^n\big)}{1 - n}\ .
\end{equation}
In the $n\to 1$ limit, we have
\begin{equation}
S_1(AB;C) = S_1(BC;A) = S_1(CA;B) = -\sum\nolimits_\alpha p_\alpha \log p_\alpha\ .
\end{equation}
Therefore, the genuine multi-entropy \eqref{eq-GME} of the state \eqref{eq-hopf-hopf} is
\begin{equation}\label{eq-212-212-1-GM}
\mathrm{GM}_1^{(3)}(A;B;C) = -\frac{1}{2} \sum\nolimits_\alpha p_\alpha \log p_\alpha\ .
\end{equation}
Note also that this genuine multi-entropy is positive and bounded from above by $\frac{1}{2}\log(k + 1)$. 

\begin{figure}[htbp]
    \centering
    \begin{subfigure}[b]{0.47\textwidth}
        \centering
        \includegraphics[width=\textwidth]{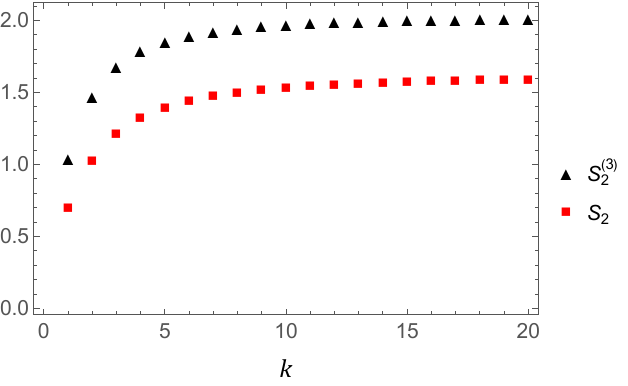}
    \end{subfigure}
    \hfill
    \begin{subfigure}[b]{0.5\textwidth}
        \centering
        \includegraphics[width=\textwidth]{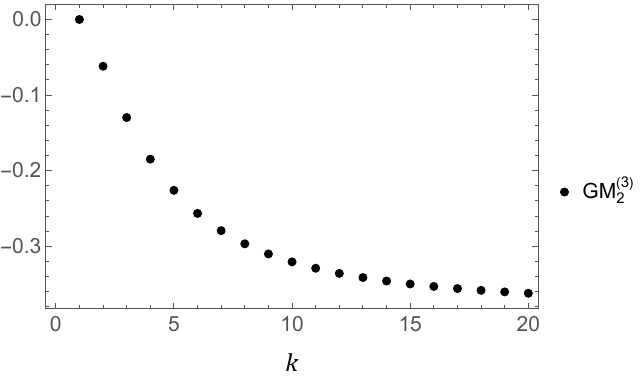}
    \end{subfigure}
    \hfill
    \caption{(Left) The second R\'enyi multi-entropy \eqref{eq-212-212-entropy} and second R\'enyi entropy \eqref{eq-633-RE2} for the link $2_1^2 + 2_1^2$ as a function of $k$. (Right) The second genuine R\'enyi multi-entropy for $2_1^2 + 2_1^2$ as a function of $k$. }
    \label{fig-212}
\end{figure}

We plot the data for the second genuine R\'enyi multi-entropy of this link in Fig.~\ref{fig-212} for $\mathrm{SU}(2)_k$ Chern-Simons theory. It can be seen that the $\mathrm{GM}_2^{(3)}$ among the three loops is always negative, indicating that the state \eqref{eq-hopf-hopf} is GHZ-like (according to \eqref{eq-GM2GHZ-le0}).
\subsubsection{\boldmath Link $6_3^3$\unboldmath}
The link $6_3^3$ (Fig.~\ref{fig-three-links} (b)) differs from the link $2_1^2 + 2_1^2$ by a Dehn twist between $B$ and $C$, thus its wavefunction is~\cite{Balasubramanian:2016sro}
\begin{equation}\label{eq-633-wave}
\ket{6_3^3} = \sum_{\alpha,\beta,\chi}\sum_m e^{-2\pi i(h_m + h_\alpha + h_\beta + h_\chi)} \frac{S_{\alpha m} S_{\beta m} S_{\chi m}}{S_{0 m}}\ket{\alpha,\beta,\chi}\ ,
\end{equation}
where $h_\alpha = \alpha(\alpha + 1)/(k + 2)$ is related to the $\mathcal{T}$-matrix through $\mathcal{T}_{\alpha\beta} = e^{2\pi i h_\alpha}\delta_{\alpha\beta}$. 
The reduced density matrix is given by
\begin{equation}\label{eq-633-rhoBC}
\begin{split}
\bra{\beta,\chi}\rho_{BC}\ket{\beta',\chi'} &= \sum_\alpha \sum_{m,n} e^{-2\pi i(h_m - h_n + h_\beta - h_{\beta'} + h_\chi - h_{\chi'})} \frac{S_{\alpha m} S_{\beta m} S_{\chi m}}{S_{0 m}}\frac{S^*_{\alpha n} S^*_{\beta' n} S^*_{\chi' n}}{S^*_{0 n}}\\
&= e^{-2\pi i(h_\beta - h_{\beta'} + h_\chi - h_{\chi'})} \sum_{m} \frac{S_{\beta m} S_{\chi m}}{S_{0 m}}\frac{S^*_{\beta' m} S^*_{\chi' m}}{S^*_{0 m}}\ .
\end{split}
\end{equation}
In the second line of \eqref{eq-633-rhoBC}, we use the property \eqref{eq-property-S}. One can see that the result in \eqref{eq-633-rhoBC} is almost identical to \eqref{eq-hopf-rhoBC}, up to a phase factor that cancels when computing the partition function on the replica space \eqref{eq-replica}. Therefore, the $n$-th R\'enyi multi-entropy among the three loops of the $6_3^3$ link is exactly the same as that in \eqref{eq-212-212-n-entropy}. We also note that the second R\'enyi entropy of the state~\eqref{eq-633-wave} coincides with \eqref{eq-633-RE2}~\cite{Balasubramanian:2016sro}.
\subsubsection{Borromean rings}
\begin{figure}[htbp]
    \centering
    \begin{subfigure}[b]{0.45\textwidth}
        \centering
        \includegraphics[width=\textwidth]{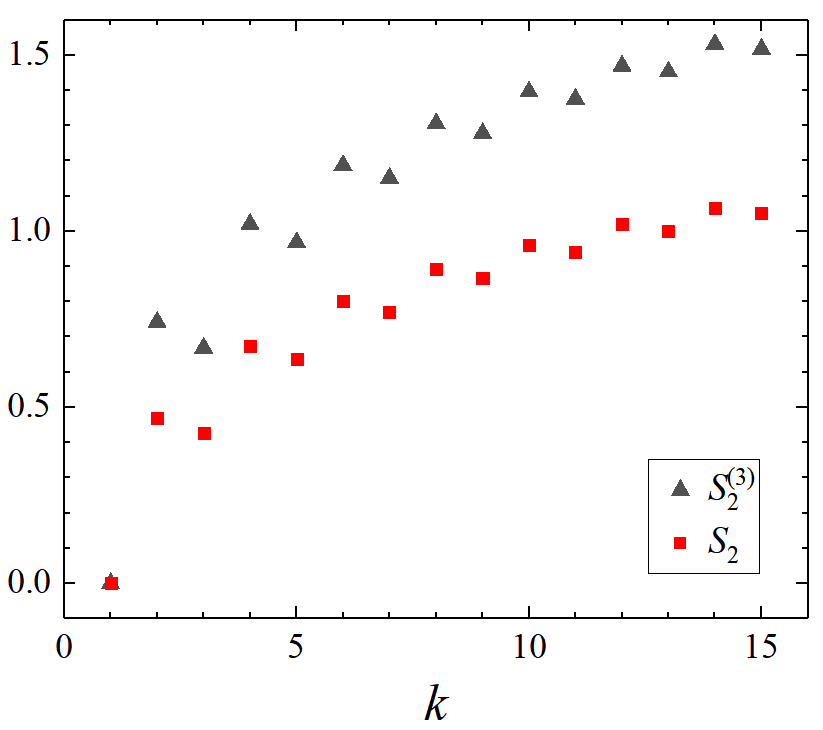}
    \end{subfigure}
    \hfill
    \begin{subfigure}[b]{0.47\textwidth}
        \centering
        \includegraphics[width=\textwidth]{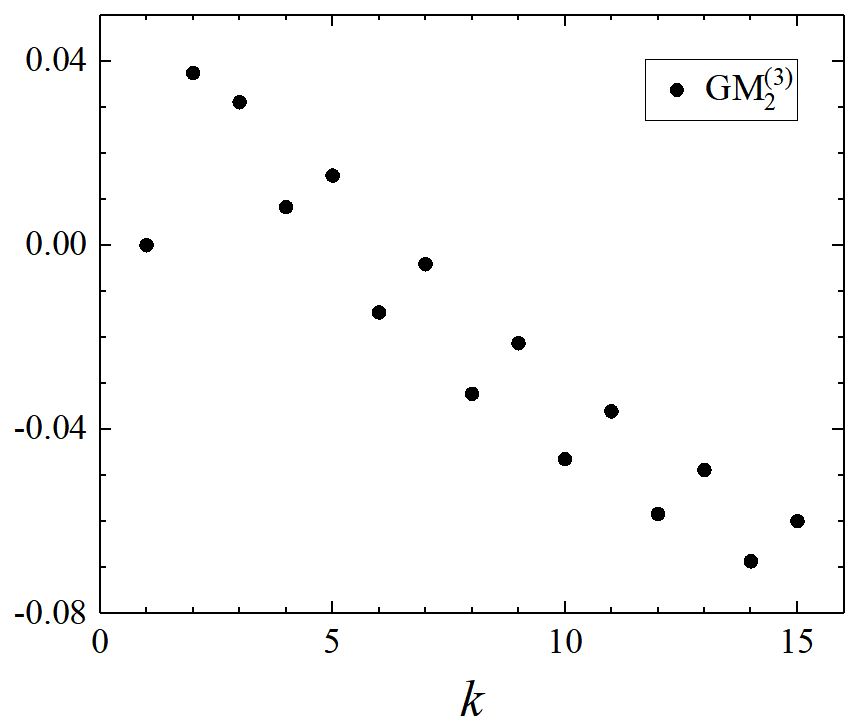}
    \end{subfigure}
    \hfill
    \caption{(Left) The second R\'enyi multi-entropy and second R\'enyi entropy for the Borromean link as a function of $k$. (Right) The second genuine R\'enyi multi-entropy for the Borromean link as a function of $k$. }
    \label{fig-Borromean}
\end{figure}

A more complicated link is the Borromean rings $6_2^3$ (Fig.~\ref{fig-three-links} (c)), whose wavefunction is given by~\cite{Balasubramanian:2016sro}
\begin{equation}
\ket{6_2^3} =\!\!\! \sum_{j = 0}^{\min(\alpha,\beta,\chi)}\!\!\!(-1)^j (q^{1/2} - q^{-1/2})^{4j} \frac{[2\alpha + j + 1]!\, [2\beta + j + 1]!\, [2\chi + j + 1]!\, ([j]!)^2}{[2\alpha - j]!\, [2\beta - j]!\, [2\chi - j]!\, ([2j + 1]!)^2}\ket{\alpha,\beta,\chi}\ ,
\end{equation}
with
\begin{equation}
[x] = \frac{q^{x/2} - q^{-x/2}}{q^{1/2} - q^{-1/2}}\ , \quad [x]! = [x] [x-1] \cdots [1]\ , \quad q = e^{2\pi i/(k + 2)}\ .
\end{equation}
For this state, we do not have an analytic formula for $S_n^{(3)}$. The numerical results for $S_2$, $S_2^{(3)}$, and $\mathrm{GM}_2^{(3)}$ are plotted in Fig.~\ref{fig-Borromean}.

\section{Logarithmic negativity for three-component links}\label{sec-nega}
Logarithmic negativity~\cite{Peres:1996dw, Vidal:2002zz} is a readily computable measure of bipartite quantum entanglement that applies to both pure and mixed states. The logarithmic negativity between subsystems $B$ and $C$ is defined as
\begin{equation}\label{eq-def-nega}
E_{\mathcal{N}}(B;C) = \log||\rho_{BC}^{\Gamma}||\ ,
\end{equation}
where $||O|| = \mathrm{Tr}\sqrt{O^\dagger O}$ is the trace norm of the matrix $O$, and $\rho_{BC}^{\Gamma}$ denotes the partial transpose of the reduced density matrix $\rho_{BC}$
\begin{equation}
\bra{\beta,\chi}\rho_{BC}^{\Gamma}\ket{\beta',\chi'} = \bra{\beta,\chi'}\rho_{BC}\ket{\beta',\chi}\ .
\end{equation}
A non-zero logarithmic negativity strictly signals the presence of pure quantum entanglement. It serves as an effective entanglement witness, consistently identifying entangled states and exclusively measuring quantum correlations, without including classical correlations.

The logarithmic negativity can also be computed using the replica trick, in close analogy to the entanglement entropy. The trace norm of the partially transposed density matrix can be expressed in terms of its eigenvalues,
\begin{equation}
\mathrm{Tr}||\rho_{BC}^{\Gamma}||=1+2 \sum_{\lambda_{i}<0}\left|\lambda_{i}\right|\ . 
\end{equation}
Therefore, one can recover the trace norm by analytically continuing the even integer powers of the partially transposed density matrix
\begin{equation}\label{eq:eigen}
\mathrm{Tr}\left(\rho_{BC}^{\Gamma}\right)^{2p}=\sum_i \lambda_i^{2p}\ , \quad p\in \mathbb{Z}_{+}
\end{equation}
to $p = 1/2$, i.e., 
\begin{equation}\label{eq:netrace}
E_{\mathcal{N}}(B;C) = \lim_{p \to 1/2} \log \mathrm{Tr}\big(\rho_{BC}^{\Gamma}\big)^{2p}\ .
\end{equation}

From \eqref{eq-rhoBC}, the partial transpose of the reduced density matrix for a three-component link state is
\begin{equation}\label{eq-rhoBCT}
\bra{\beta,\chi}\rho_{BC}^{\Gamma}\ket{\beta',\chi'} =\frac{1}{k^2} e^{\frac{2\pi i}{k}(\beta \chi' - \beta' \chi)L_{BC}} \eta\!\left[(\beta-\beta')L_{AB}-(\chi-\chi')L_{CA}\right]\!\ .
\end{equation}
When $p = 1$, we have
\begin{equation}\label{eq-EN-replica2}
\begin{split}
\mathrm{Tr}\big(\rho_{BC}^{\Gamma}\big)^{2} &=\! \sum_{\beta_{1,2}, \chi_{1,2}}\! \frac{1}{k^4} \eta\!\left[(\beta_1\!-\!\beta_2)L_{AB} - (\chi_1\!-\!\chi_2)L_{CA}\right]\, \eta\!\left[(\beta_2\!-\!\beta_1)L_{AB} - (\chi_2\!-\!\chi_1)L_{CA}\right]\\
&= \frac{1}{k^4}\times k^2 \times k\gcd(k,L_{CA},L_{AB})\\
&= \frac{1}{k} \gcd(k,L_{CA},L_{AB})\ ,
\end{split}
\end{equation}
where in the second line, the factor $k\gcd(k, L_{CA}, L_{AB})$ arises from the fact that there are $k\gcd(k, L_{CA}, L_{AB})$ sets of $\{\beta_1 - \beta_2, \chi_1 - \chi_2\}$ satisfying the $\eta$-constraint, while the factor $k^2$ comes from the two free variables. For $p > 1$, it is difficult to perform a similar calculation. Nevertheless, we can first obtain the result for $L_{CA} = 0$ (see appendix~\ref{appx-negativity} for the derivation) 
\begin{equation}\label{eq-EN-replica2p}
\mathrm{Tr}\big(\rho_{BC}^{\Gamma}\big)^{2p}\Big|_{L_{CA} = 0} = \frac{1}{k^{2p - 1}}\gcd(k,L_{AB},L_{BC})^{2p - 2}\gcd(k,L_{AB})\ .
\end{equation}
A complete result with nontrivial $L_{CA}$ should reduce to \eqref{eq-EN-replica2} when $p = 1$, and should be symmetric between $B$ and $C$. Therefore, we make the following improvement
\begin{align}
&\gcd(k,L_{AB}) = \gcd(k,0,L_{AB}) \to \gcd(k,L_{CA},L_{AB})\ ,\\
&\gcd(k,L_{AB},L_{BC}) = \gcd(k,0,L_{AB},L_{BC}) \to \gcd(k,L_{CA},L_{AB},L_{BC})
\end{align}
in \eqref{eq-EN-replica2p}, and we eventually obtain
\begin{equation}\label{eq-tracenorm-2p}
\mathrm{Tr}\big(\rho_{BC}^{\Gamma}\big)^{2p} = \frac{1}{k^{2p - 1}}\gcd(k,L_{CA},L_{AB},L_{BC})^{2p - 2}\gcd(k,L_{CA},L_{AB})\ .
\end{equation}
Therefore, the logarithmic negativity for the 3-link state \eqref{eq-wave3} is given by\footnote{See the left panel of Fig.~\ref{fig-Venn-enatnglement} for the Venn diagram corresponding to the logarithmic negativity~\eqref{eq-negativity-3link}.}
\begin{equation}\label{eq-negativity-3link}
E_{\mathcal{N}}(B;C) = \lim_{p\to 1/2} \log \mathrm{Tr}\big(\rho_{BC}^{\Gamma}\big)^{2p} = \log\frac{\gcd(k,L_{CA},L_{AB})}{\gcd(k,L_{CA},L_{AB},L_{BC})}\ .
\end{equation}
One can see that it reduces to the entanglement entropy between $B$ and $C$ when $L_{CA} = L_{AB} = 0$. We note again that the expressions \eqref{eq-tracenorm-2p} and \eqref{eq-negativity-3link} are conjectures, motivated by the rigorous results \eqref{eq-EN-replica2} and \eqref{eq-EN-replica2p}, and further supported by extensive numerical tests.\footnote{We have numerically verified \eqref{eq-negativity-3link} for all three-component links with $k \le 18$, and also tested \eqref{eq-tracenorm-2p} with $p = 2$ for the same range of $k$.}
\subsection{Third-order negativity for three-links}\label{sec-3log-nega}
For future reference, we also compute the third-order negativity for three-component links. It is given by\footnote{There is no contradiction between \eqref{eq-third-negativity} and \eqref{eq-tracenorm-2p}, since the latter is valid only for integer $p$. We also note that our definition of the third-order negativity in \eqref{eq-third-negativity} includes an additional factor of $1/2$.}
\begin{equation}\label{eq-third-negativity}
\mathcal{E}_3(B;C):= -\frac{1}{2}\log\mathrm{Tr}\big(\rho_{BC}^{\Gamma})^3 = \log\frac{k}{\gcd(k,L_{CA}, L_{AB}, L_{BC})}\ .
\end{equation}
See appendix~\ref{appx-nega3} for a non-rigorous proof of this formula. We have also numerically verified \eqref{eq-third-negativity} for all three-component links with $k \le 18$.  
\section{\boldmath Entanglement structure of links\unboldmath}\label{sec-relations}
The link states in Abelian Chern-Simons theory, given by \eqref{eq-waveN}, are stabilizer states in the context of quantum information theory \cite{Balasubramanian:2018por}. In this section, we employ the decomposition properties \eqref{eq-decomposition} and \eqref{eq-general-decompose} of stabilizer states to explore the physical meanings of $\mathrm{GM}_1^{(3)}$ and $E_{\mathcal{N}}$ within this class of states. From these decomposition properties, we derive two relations connecting entanglement entropy, third-order negativity, genuine tri-entropy, and logarithmic negativity \eqref{eq-relation-m}. Using these relations, we verify the consistency of our results \eqref{eq-n-GME-3link}, \eqref{eq-negativity-3link}, and \eqref{eq-third-negativity}. Finally, we propose possible upper and lower bounds for the genuine tri-entropy in general finite-dimensional systems in Section~\ref{subsec-bound-GME-Nlinks}.
\subsection{Prime-power qudit stabilizer states}\label{subsubsec-p-pl}
In this section we consider $N$-qudit stabilizer states where each qudit has dimension $k = p^\epsilon$ ($\epsilon$ is positive integer). We refer to such qudits as power-prime qudits.  Any tripartite power-prime qudit stabilizer state $\ket{\mathtt{S}}_{ABC}$ is locally equivalent to a tensor product of tripartite GHZ$_p$-states $\ket{\mathtt{GHZ}_p}_{ABC} \propto \sum_{i = 0}^{p - 1}\ket{iii}$, bipartite maximally entangled EPR$_p$-pairs $\ket{\mathtt{EPR}_p}_{AB} \propto \sum_{i = 0}^{p - 1}\ket{ii}$, and unentangled single qudit states $\ket{+}$, that is, there exists a local unitary $U = U_A\otimes U_B\otimes U_C$ such that~\cite{Wong:2025cwc}\footnote{See~\cite{Bravyi05, Looi:2011jrm} for previous study on qubit and squarefree qudit case. }
\begin{equation}\label{eq-decomposition}
\begin{split}
U_A\otimes U_B\otimes U_C\ket{\mathtt{S}_{p^\epsilon}}_{ABC} =&\, \ket{\mathtt{GHZ}_p}_{ABC}^{\otimes m_{ABC}}\\
&\otimes \ket{\mathtt{EPR}_p}_{AB}^{\otimes m_{AB}}\otimes \ket{\mathtt{EPR}_p}_{BC}^{\otimes m_{BC}}\otimes \ket{\mathtt{EPR}_p}_{CA}^{\otimes m_{CA}}\\
&\otimes \ket{+}_{A}^{\otimes m_{A}}\otimes \ket{+}_{B}^{\otimes m_{B}}\otimes \ket{+}_{C}^{\otimes m_{C}}\ ,
\end{split}
\end{equation}
where $m_{ABC}, m_{AB}, \dots, m_{A}, \dots$ are non-negative integers.
For the product state in \eqref{eq-decomposition}, we have
\begin{equation}
\begin{split}
\mathrm{GM}_n^{(3)}[\mathtt{S}_{ABC}] =&\, m_{ABC}\, \mathrm{GM}_n^{(3)}[\mathtt{GHZ}_{ABC}]\\ 
&\,+ m_{AB}\, \mathrm{GM}_n^{(3)}[\mathtt{EPR}_{AB}] + m_{BC}\, \mathrm{GM}_n^{(3)}[\mathtt{EPR}_{BC}] + m_{CA}\, \mathrm{GM}_n^{(3)}[\mathtt{EPR}_{CA}]\\
&\,+ m_A\, \mathrm{GM}_n^{(3)}[+_A] + m_B\, \mathrm{GM}_n^{(3)}[+_B] + m_C\, \mathrm{GM}_n^{(3)}[+_C]\\
=&\, m_{ABC}\times\Big(\frac{1}{n} - \frac{1}{2}\Big)\log p\ ,
\end{split}
\end{equation}
where we have used the property that the genuine R\'enyi multi-entropy vanishes for product states of the form~\eqref{eq-GM-0}. Therefore, for stabilizer states, the genuine tri-entropy precisely extracts the number of GHZ$_p$-states.

Similarly, we consider the logarithmic negativity of Eq.~\eqref{eq-decomposition}, where on the right-hand side we use $E_{\mathcal{N}}^{B;C}$ as a shorthand for $E_{\mathcal{N}}(B;C)$:
\begin{equation}
\begin{split}
E_{\mathcal{N}}(B;C)[\mathtt{S}_{ABC}] =&\, m_{ABC}\, E_{\mathcal{N}}^{B;C}[\mathtt{GHZ}_{ABC}]\\ 
&\,+ m_{AB}\, E_{\mathcal{N}}^{B;C}[\mathtt{EPR}_{AB}] + m_{BC}\, E_{\mathcal{N}}^{B;C}[\mathtt{EPR}_{BC}] + m_{CA}\, E_{\mathcal{N}}^{B;C}[\mathtt{EPR}_{CA}]\\
=&\, m_{BC}\log p\ ,
\end{split}
\end{equation}
where we have used the following facts,
\begin{equation}
E_{\mathcal{N}}^{B;C}[\mathtt{GHZ}_{ABC}] = E_{\mathcal{N}}^{B;C}[\mathtt{EPR}_{AB}] = E_{\mathcal{N}}^{B;C}[\mathtt{EPR}_{CA}] = 0\ , \quad E_{\mathcal{N}}^{B;C}[\mathtt{EPR}_{BC}] = \log p\ .
\end{equation}
Therefore, the logarithmic negativity captures the number of EPR-pairs in a stabilizer state. We summarize these two results as follows: for power-prime qudit stabilizer states, 
\begin{align}
\label{eq-GHZ-Count}
2\, \mathrm{GM}_1^{(3)}[\mathtt{S}_{ABC}] &= m_{ABC} \log p\ ,\\
\label{eq-EPR-Count}
E_{\mathcal{N}}(B;C)[\mathtt{S}_{ABC}] &= m_{BC} \log p\ .
\end{align}

We can verify these relations using link states. The number of GHZ-states, $m_{ABC}$, in a stabilizer state can be determined by the following quantity~\cite{Nezami:2016zni, Salton:2016qpp}\footnote{The quantity $\mathrm{Tr} (\rho_{BC}^{\Gamma})^3$ is one of the six algebraically independent local invariants of pure 3-qubit states~\cite{Anthony0001116}, 
\begin{equation*}
I_1 = \braket{\Psi|\Psi}\ , \quad I_2 = \mathrm{Tr}(\rho_C)^2\ , \quad I_3 = \mathrm{Tr}(\rho_B)^2\ , \quad I_4 = \mathrm{Tr}(\rho_A)^2\ , \quad I_5 = \mathrm{Tr}(\rho_{BC}^\Gamma)^3\ , \quad I_6 = \tau_{ABC}\ ,
\end{equation*}
where $\tau_{ABC}$ is the 3-tangle~\cite{Coffman:1999jd}. We thank Chen-Te Ma for discussions.},
\begin{equation}\label{eq-number-GHZ}
m_{ABC} \log p = S(A) + S(B) + S(C) + \log\mathrm{Tr}\big(\rho_{BC}^{\Gamma}\big)^3\ ,
\end{equation}
where the last term is the third-order negativity discussed in Section~\ref{sec-3log-nega}. For three-component link states, it is given explicitly by Eq.~\eqref{eq-third-negativity}. Substituting \eqref{eq-3-RE2} and \eqref{eq-third-negativity} into \eqref{eq-number-GHZ}, we obtain
\begin{equation}\label{eq-GHZ-number-new}
m_{ABC}\log p = \log\frac{k\cdot \gcd(k,L_{CA},L_{AB},L_{BC})^2}{\gcd(k,L_{CA},L_{AB}) \gcd(k,L_{AB},L_{BC}) \gcd(k,L_{BC},L_{CA})}\ ,
\end{equation}
which is nothing but twice the genuine tri-entropy of three-link states~\eqref{eq-n-GME-3link}. 
It follows from Eq.~\eqref{eq-GHZ-number-new} that $m_{ABC}$ must be an integer when $k = p^\epsilon$.

The number of EPR-pairs (eg. $m_{BC}$, $m_{CA}$ and $m_{AB}$) in \eqref{eq-decomposition} can be extracted from the mutual information \cite{Nezami:2016zni}, 
\begin{equation}\label{eq-number-EPR}
(2m_{BC} + m_{ABC}) \log p = I(B;C) = S(B) + S(C) - S(BC)\ .
\end{equation}
Using \eqref{eq-3-RE2} and \eqref{eq-GHZ-number-new}, we obtain
\begin{equation}\label{eq-EPR-pairs}
m_{BC} \log p = \log\frac{ \gcd(k,L_{CA},L_{AB})}{\gcd(k,L_{CA},L_{AB},L_{BC})}\ ,
\end{equation}
which is exactly the logarithmic negativity between $B$ and $C$ \eqref{eq-negativity-3link}. One can see that the $m_{BC}$ in \eqref{eq-EPR-pairs} is an integer if $k = p^\epsilon$. 
\subsection{General qudit stabilizer states}\label{subsubsec-general-k}

Now we turn to more general $N$-qudit tripartite stabilizer states $\ket{\mathtt{S}}_{ABC}$, where the dimension $k$ of each qudit is not necessarily a prime-power number. Every integer $k$ admits a prime factorization,
\begin{equation}\label{eq-general-k}
k = p_1^{\epsilon_1} p_2^{\epsilon_2} \cdots p_\ell^{\epsilon_\ell}\ .
\end{equation}
There always exists some single-qudit unitary transformations $U_i$ such that~\cite{Looi:2011jrm}
\begin{equation}\label{eq-general-decompose}
U_1\otimes U_2\otimes \cdots \otimes U_N \ket{\mathtt{S}}_{ABC} = \bigotimes_{i=1}^{\ell} \ket{\mathtt{S}_i}\ ,
\end{equation}
where $\ket{\mathtt{S}_i}$ is a stabilizer state on $N$ qudits of dimension $p_i^{\epsilon_i}$, which can be further decomposed as in Eq.~\eqref{eq-decomposition}.
It is then straightforward to see that for general $k$ given by Eq.~\eqref{eq-general-k}, Eqs.~\eqref{eq-GHZ-Count} and~\eqref{eq-EPR-Count} generalize to
\begin{align}
\label{eq-k-GHZ}
2\,\mathrm{GM}_1^{(3)}[\mathtt{S}_{ABC}] &= \sum\nolimits_{i=1}^{\ell} m_{ABC,i}\, \mathrm{log}\ p_i\ ,\\
E_{\mathcal{N}}(B;C)[\mathtt{S}_{ABC}] &= \sum\nolimits_{i=1}^{\ell} m_{BC,i}\, \mathrm{log}\ p_i\ ,
\end{align}
where $m_{ABC,i}$ and $m_{BC,i}$ denote the numbers of $\ket{\mathtt{GHZ}_{p_i}}_{ABC}$ and $\ket{\mathtt{EPR}_{p_i}}_{BC}$ appearing in the decomposition, respectively.
We can thus conclude that for general stabilizer states (with arbitrary qudit dimension), the genuine tri-entropy quantifies the tripartite entanglement generated by GHZ-states, while the logarithmic negativity between $B$ and $C$ quantifies the bipartite entanglement generated by EPR-pairs between $B$ and $C$.

Here we use $\ket{\mathrm{GHZ}_6}$ as a simple example to illustrate \eqref{eq-k-GHZ}. The state $\ket{\mathrm{GHZ}_6}$ can be decomposed as
\begin{equation}
\begin{split}
\ket{\mathrm{GHZ}_6} 
&= \frac{1}{\sqrt{6}}\sum_{i = 0}^{5} \ket{iii} = \frac{1}{\sqrt{6}}\sum_{i = 00,01,02,10,11,12}\ket{iii}\\
&= \frac{1}{\sqrt{6}}\big(\ket{0{\color{blue}0}0{\color{blue}0}0{\color{blue}0}} + \ket{0{\color{blue}1}0{\color{blue}1}0{\color{blue}1}} + \ket{0{\color{blue}2}0{\color{blue}2}0{\color{blue}2}} + \ket{1{\color{blue}0}1{\color{blue}0}1{\color{blue}0}} + \ket{1{\color{blue}1}1{\color{blue}1}1{\color{blue}1}} + \ket{1{\color{blue}2}1{\color{blue}2}1{\color{blue}2}}\big)\\
&= \frac{1}{\sqrt{2}}\big(\ket{000} + \ket{111}\big)\otimes \frac{1}{\sqrt{3}}\big(\ket{{\color{blue}000}} + \ket{{\color{blue}111}} + \ket{{\color{blue}222}}\big)\\
&= \ket{\mathtt{GHZ}_2}\otimes \ket{\mathtt{GHZ}_3}\ ,
\end{split}
\end{equation}
from which we can read that $m_{ABC,1} = m_{ABC,2} = 1$ (this is also consistent with $6 = 2\times 3$). The genuine tri-entropy of the state $\ket{\mathrm{GHZ}_6}$ is
\begin{equation}
2\mathrm{GM}_1^{(3)}[\mathrm{GHZ}_6] = \log 6 = \log 2 + \log 3\ ,
\end{equation}
which is consistent with \eqref{eq-k-GHZ}.

Using the same method, we also find that for stabilizer states, the second genuine R\'enyi tri-entropy vanishes:
\begin{equation}\label{eq-stabilizer-GM2}
\mathrm{GM}_2^{(3)}[\mathtt{S}_{ABC}] = 0\ ,
\end{equation}
since the second genuine R\'enyi tri-entropy is zero for GHZ-states. Eq.~\eqref{eq-stabilizer-GM2} is consistent with our $N$-link results in Section~\ref{sec-Nlink}.

For general $N$-qudit stabilizer states with arbitrary qudit dimension, although we cannot isolate a specific $m_{ABC,i}$ using $\mathrm{GM}_1^{(3)}$, the following two relations hold:\footnote{These two relations follow directly from the decomposition properties \eqref{eq-decomposition} and \eqref{eq-general-decompose} of stabilizer states. We check \eqref{eq-relation-m} for all four-links $\mathcal{L}^{2,1,1}$ with $k\le 7$ (here by ``all four-links $\mathcal{L}^{2,1,1}$'' we mean that all independent sets of $(L_{A_1 B}, L_{A_2 B}, L_{BC}, L_{C A_1}, L_{C A_2})\in \mathbb{Z}^{\otimes 5}$), for all five-links $\mathcal{L}^{3,1,1}$ with $k\le 5$, for all five-links $\mathcal{L}^{2,2,1}$ with $k\le 4$, for all six-links $\mathcal{L}^{4,1,1}$ with $k\le 3$, for all six-links $\mathcal{L}^{2,2,2}$ with $k\le 2$. }
\begin{equation}\label{eq-relation-m}
\begin{split}
S(A) + S(B) + S(C) + \log\mathrm{Tr}\big(\rho_{BC}^{\Gamma}\big)^3 &= 2\,\mathrm{GM}_1^{(3)}(A;B;C)\ ,\\
-\frac{1}{2}\log\mathrm{Tr}\big(\rho_{BC}^{\Gamma}\big)^3 - S(A) &= E_{\mathcal{N}}(B;C)\ .
\end{split}
\end{equation}
\subsection{Possible lower and upper bounds of the genuine tri-entropy}\label{subsec-bound-GME-Nlinks}
From the decomposition properties \eqref{eq-decomposition} and \eqref{eq-general-decompose} of stabilizer states, together with the fact \eqref{eq-k-GHZ} that the genuine tri-entropy counts the total tripartite entanglement generated by GHZ-states, we can see that for stabilizer states, the genuine tri-entropy is bounded by
\begin{equation}\label{eq-bound}
0 \le \mathrm{GM}_1^{(3)}(A;B;C) \le \frac{1}{2}\log\! \Big(\!\min\!\left\{\dim \mathcal{H}_A, \dim \mathcal{H}_B, \dim \mathcal{H}_C\right\}\!\Big)\ ,
\end{equation}
where $\dim \mathcal{H}_A$ is the dimension of the Hilbert space associated with the subsystem $A$. 
The bound \eqref{eq-bound} is reminiscent of the bound of entanglement entropy\footnote{By definition the entanglement entropy is the $\mathtt{q} = 2$ multi-entropy, and according to the construction of genuine multi-entropy (Section~\ref{sec-GME-def}), it is also the $\mathtt{q} = 2$ genuine multi-entropy. }, 
\begin{equation}
0\le S^{(2)}(A;B) = \mathrm{GM}_1^{(2)}(A;B)\le \log\! \Big(\!\min\!\left\{\dim \mathcal{H}_A, \dim \mathcal{H}_B\right\}\!\Big)\ ,
\end{equation}
which holds for general finite systems (instead of just stabilizer states). 
Therefore, it is temping to conjecture that the bound \eqref{eq-bound} of $\mathrm{GM}_1^{(3)}(A;B;C)$ also holds for \textit{general finite systems}. We now check several examples. The bound \eqref{eq-bound} is satisfied by the link $2_1^2 + 2_1^2$ (and the link $6_3^3$) in $\mathrm{SU}(2)_k$ Chern-Simons theory \eqref{eq-212-212-1-GM}\footnote{For states living on the toric boundary in $\mathrm{SU}(2)_k$ Cherm-Simons theory, the dimension of the associated Hilbert is $\dim \mathcal{H}\big(\mathbb{T}^2;\mathrm{SU}(2)_k\big) = k + 1$. }, 
\begin{equation}
0\le \mathrm{GM}_1^{(3)}\big[2_1^2 + 2_1^2, \mathrm{SU}(2)_k\big] = \mathrm{GM}_1^{(3)}\big[6_3^3, \mathrm{SU}(2)_k\big] \le \frac{1}{2}\log (k + 1)\ .
\end{equation}
For the generalized GHZ-state \eqref{eq-gGHZ}, 
\begin{equation}
0 \le \mathrm{GM}_1^{(3)}[\widetilde{\text{GHZ}}_k] = -\frac{1}{2}\sum |\lambda_i|^2 \log |\lambda_i|^2 \le \frac{1}{2}\log k\ ,
\end{equation}
where the equality holds for the $\text{GHZ}_k$-state \eqref{eq-GHZ}. For the state $\psi^{N_A, N_B, N_C}$ in the (1+1)d Chern-Simons theory, its $\mathrm{GM}_1^{(3)}$ is given by~\eqref{eq-2d-GM1} and we have 
\begin{equation}
0\le \mathrm{GM}_1^{(3)}\big[\psi^{N_A,N_B,N_C}_{m}\big]\le \frac{1}{2}\log|G|\ ,
\end{equation}
where the upper bound is reached when $G$ is an abelian group. See appendix~\ref{sec-1+1d} for details. 

In \cite{Iizuka:2025ioc}, the authors conjecture that for the W-state $\ket{\text{W}} = (\ket{001} + \ket{010} + \ket{100})/\sqrt{3}$, 
\begin{equation}
S_1^{(3)}[\text{W}] = c_{\text{W}}\log 3\ , \quad \mathrm{GM}_1^{(3)}[\text{W}] = \log \frac{2}{3^{3/2 - c_{\text{W}}}}\ ,
\end{equation}
where $c_{\text{W}}$ is a numerical constant greater than 1. The $n = 2,\dots,6$ R\'enyi multi-entropies of W-state are~\cite{Iizuka:2025ioc}
\begin{equation}\label{eq-W-RME}
\begin{split}
&S_2^{(3)}[\text{W}] = \log 3\ , \quad S_3^{(3)}[\text{W}] \approx 0.93\log 3\ , \quad S_4^{(3)}[\text{W}] \approx 0.88 \log 3\ , \\
&S_5^{(3)}[\text{W}] \approx 0.84 \log 3\ , \quad S_6^{(3)}[\text{W}] \approx 0.82\log 3\ .
\end{split}
\end{equation}
If one plots $n$ along the $x$-axis and the prefactor of $\log 3$ along the $y$-axis, the data points exhibit a convex ($\cup$-shaped) functional behavior, suggesting $c_{\text{W}} > 1.07$.
Alternatively, plotting $1/n^2$ versus the same prefactor gives a concave ($\cap$-shaped) trend, implying $c_{\text{W}} < 1.38$.
In summary, $c_{\text{W}}$ is likely in the range $1.07 \lesssim c_{\text{W}} \lesssim 1.38$.
Although it remains uncertain whether the W-state exactly satisfies the bound \eqref{eq-bound}, the data in \eqref{eq-W-RME} strongly suggest that it most likely does.
\section{Conclusion and discussion}

In this work, we have studied the multipartite entanglement structure of link states in three-dimensional Chern-Simons theory. For three-component links, we obtained analytic expressions for their Rényi multi-entropy and Rényi entanglement negativity. For $N$-component links, we found that the logarithmic negativity counts the bipartite entanglement generated by EPR-pairs, while the genuine multi-entropy counts tripartite entanglement generated by GHZ-states. 

There are several directions for future investigation. First, it would be interesting to generalize the analytic expression of $S_n^{(3)}$ in Eq.~\eqref{eq-n-multiE-3link} to the quadripartite or general $N$-component case. Second, one may attempt to prove the bounds in Eq.~\eqref{eq-bound} for general finite-dimensional systems. Finally, following the ideas of Refs.~\cite{Hung:2018rhg, Zhou:2019ezk}, it would be worthwhile to explore possible connections between multipartite entanglement and quantum anomalies in topological field theories.
We expect that our results will shed light on the structure of multipartite entanglement in topological quantum field theories and its potential connections to holography and quantum gravity.
\acknowledgments
We would like to thank Jinwei Chu and Chen-Te Ma for useful discussions. This work is supported by NSFC grant 12375063 and also supported by Shanghai Oriental
Talents Program. YZ is also supported by NSFC 12247103 through Peng Huanwu Center for Fundamental Theory.
\appendix
\section{Second genuine R\'enyi tri-entropy of generalized GHZ-states \eqref{eq-gGHZ}}\label{appx-GHZ}
The second R\'enyi tri-entropy of the generalized GHZ-state \eqref{eq-gGHZ} is~\cite{Gadde:2022cqi}\footnote{We use a slightly different definition of the R\'enyi multi-entropy in \eqref{eq-RME-def} compared to \cite{Gadde:2022cqi}, thus \eqref{eq-gGHZ-2ME} has an extra $1/2$ factor. }
\begin{equation}\label{eq-gGHZ-2ME}
S_2^{(3)}[\widetilde{\text{GHZ}}_k^N] = -\frac{1}{2}\log\Big(\sum\nolimits_i |\lambda_i|^8\Big)\ ,
\end{equation}
and the second R\'enyi entropy is
\begin{equation}
S_2[\widetilde{\text{GHZ}}_k^N] = -\log\Big(\sum\nolimits_i |\lambda_i|^4\Big)\ .
\end{equation}
Therefore, we have
\begin{equation}\label{eq-GM2-gGHZ}
\mathrm{GM}_2^{(3)}[\widetilde{\text{GHZ}}_k^N] = S_2^{(3)} - \frac{3}{2}S_2 = \frac{1}{2}\log\frac{\big(\sum_i |\lambda_i|^4 \big)^3}{\sum_i |\lambda_i|^8} = \frac{1}{2}\log\frac{\big(\sum_i p_i^2\big)^3}{\sum_i p_i^4}\ ,
\end{equation}
where at the last equal sign we introduce $p_i = |\lambda_i|^2$ and $\{p_i\}$ is a normalized probability distribution since we consider normalized state \eqref{eq-gGHZ}. In this appendix we will prove that the second genuine R\'enyi tri-entropy of the generalized GHZ-state \eqref{eq-GM2-gGHZ} is nonpositive, i.e., 
\begin{equation}\label{eq-gGHZ-toProof}
\Big(\sum\nolimits_i p_i^2\Big)^3 \le \sum\nolimits_i p_i^4\ .
\end{equation}
For convenience, we introduce the following notation
\begin{equation}
P_m := \sum\nolimits_i p_i^m\ .
\end{equation}

\noindent\underline{Proof of \eqref{eq-gGHZ-toProof}.} 
\begin{enumerate}[Step 1:]
\item Prove $(P_2)^2 \le P_3$. Consider a discrete random variable $X$ which takes values $p_i$ with probabilities $p_i$. According to the non-negativity of variance, we have
\begin{equation}
\mathrm{Var}(X) = \mathrm{E}[X^2] - \mathrm{E}[X]^2 =\!\Big(\sum\nolimits_i p_i^2\cdot p_i\Big) - \Big(\sum\nolimits_i p_i\cdot p_i\Big)^2\!= P_3 - (P_2)^2 \ge 0\ .
\end{equation}

\item Prove $(P_3)^2 \le P_2 P_4$. This follows directly from the Cauchy-Schwarz inequality
\begin{equation}
\Big(\sum\nolimits_i u_i v_i\Big)^2 \le \Big(\sum\nolimits_i u_i^2\Big) \Big(\sum\nolimits_i v_i^2\Big)
\end{equation}
by taking $u_i = p_i^2$ and $v_i = p_i$. 
\end{enumerate}
From Step 1 and 2, we obtain
\begin{equation}
(P_2)^4 \le (P_3)^2 \le P_2 P_4\ ,
\end{equation}
and since $P_2 > 0$, we end up with $(P_2)^3 \le P_4$, which is exactly \eqref{eq-gGHZ-toProof}.

\section{Derivation of \eqref{eq-improve}}\label{appx-derive}
We take $n = 3$ as an example. When $L_{BC} = 0$, the partition function on the replica space is a summation over $\beta_{1,\dots,n^2 = 9}, \chi_{1,\dots,n^2 = 9}$ of $1/k^{2n^2} = 1/k^{18}$ with the following $n^2 = 9$ constraints
\begin{equation}\label{eq-9-constrains}
\begin{split}
1 = \quad &\eta\big[\beta_{14}L_{AB}+\chi_{12}L_{CA}\big]\, \eta\big[\beta_{25}L_{AB}+\chi_{23}L_{CA}\big]\, \eta\big[\beta_{36}L_{AB}+\chi_{31}L_{CA}\big]\\
\times &\eta\big[\beta_{47}L_{AB}+\chi_{45}L_{CA}\big]\, \eta\big[\beta_{58}L_{AB}+\chi_{56}L_{CA}\big]\, \eta\big[\beta_{69}L_{AB}+\chi_{64}L_{CA}\big]\\
\times &\eta\big[\beta_{71}L_{AB}+\chi_{78}L_{CA}\big]\, \eta\big[\beta_{82}L_{AB}+\chi_{89}L_{CA}\big]\, \eta\big[\beta_{93}L_{AB}+\chi_{97}L_{CA}\big]\ ,\\
\end{split}
\end{equation}
with the notation 
\begin{equation}\label{eq-notation-ij}
\beta_{ij}:= (\beta_i - \beta_j)\!\!\!\! \mod k\ , \quad \chi_{ij}:= (\chi_i - \chi_j)\!\!\!\! \mod k\ .
\end{equation}
Using the property \eqref{eq-property} to the first row in \eqref{eq-9-constrains}, we obtain
\begin{equation}
\begin{split}
&\,\eta\big[\beta_{14}L_{AB}+\chi_{12}L_{CA}\big]\, \eta\big[\beta_{25}L_{AB}+\chi_{23}L_{CA}\big]\, {\color{blue}\eta\big[\beta_{36}L_{AB}+\chi_{31}L_{CA}\big]}\\
=&\, \eta\big[\beta_{14}L_{AB}+\chi_{12}L_{CA}\big]\, \eta\big[\beta_{25}L_{AB}+\chi_{23}L_{CA}\big]\, {\color{blue}\eta\big[(\beta_{14} + \beta_{25} + \beta_{36})L_{AB}\big]}\ .
\end{split}
\end{equation}
Similarly, applying the property to every row and every column, we get
\begin{equation}
\begin{split}
1 = \quad&\eta\big[\beta_{14}L_{AB}+\chi_{12}L_{CA}\big]\qquad\!\!\eta\big[\beta_{25}L_{AB}+\chi_{23}L_{CA}\big]\qquad\!{\color{blue}\eta\big[(\beta_{14} + \beta_{25} + \beta_{36})L_{AB}\big]}\\
\times &\eta\big[\beta_{47}L_{AB}+\chi_{45}L_{CA}\big]\qquad\!\!\eta\big[\beta_{58}L_{AB}+\chi_{56}L_{CA}\big]\qquad\!{\color{blue}\eta\big[(\beta_{47} + \beta_{58} + \beta_{69})L_{AB}\big]}\\
\times &{\color{red}\eta\big[(\chi_{12} + \chi_{45} + \chi_{78})L_{CA}\big]}\,{\color{red}\eta\big[(\chi_{23} + \chi_{56} + \chi_{89})L_{CA}\big]}\, \eta\big[0\big]\ .
\end{split}
\end{equation}
Now we can count the number of solutions. There are 
\begin{enumerate}[\textbullet]
\item $k\cdot\gcd(k,L_{CA},L_{AB})$ possible $\{\beta_{14},\chi_{12}\}\in \mathbb{Z}_k^{\otimes 2}$ satisfying $1 = \eta\big[\beta_{14}L_{AB} + \chi_{12}L_{CA}\big]$. Similar for $\{\beta_{25}, \chi_{23}\}$, $\{\beta_{47}, \chi_{45}\}$, and $\{\beta_{58}, \chi_{56}\}$. 
\item $\gcd(k,L_{AB})$ possible ${\color{blue}\beta_{36}}\in \mathbb{Z}_k$ satisfying $1 = \eta\big[(\beta_{14} + \beta_{25} + \beta_{36})L_{AB}\big]$ (given $\beta_{14}$ and $\beta_{25}$). Similar for ${\color{blue}\beta_{69}}$. 
\item $\gcd(k,L_{CA})$ possible ${\color{red}\chi_{78}}\in \mathbb{Z}_k$ satisfying $1 = \eta\big[(\chi_{12} + \chi_{45} + \chi_{78})L_{CA}\big]$ (given $\chi_{12}$ and $\chi_{45}$). Similar for ${\color{red}\chi_{89}}$. 
\item Note that there are still $2n^2 - 2(n - 1)^2 - {\color{blue}(n - 1)} - {\color{red}(n - 1)} = 2n$ free variables, each of them can take $k$ values. 
\end{enumerate}
Therefore, 
\begin{equation}
\begin{split}
\mathcal{Z}_{n^2}^{(3)} &= \frac{1}{k^{2n^2}}\big(k\cdot\gcd(k,L_{CA},L_{AB})\big)^{(n - 1)^2}\!\times{\color{blue}\gcd(k,L_{AB})^{n - 1}}\times{\color{red}\gcd(k,L_{CA})^{n - 1}}\times k^{2n}\\
&= \frac{k^{1-n^2}}{\gcd(k,L_{CA},L_{AB})^{-(1 - n)^2}\times\gcd(k,L_{AB})^{1 - n}\times\gcd(k,L_{CA})^{1 - n}}\ ,
\end{split}
\end{equation}
from which we can get \eqref{eq-improve} using the definition of $S_n^{(3)}$. 
\section{Derivation of \eqref{eq-EN-replica2p}}\label{appx-negativity}
We take $p = 3$ as an example. When $L_{CA} = 0$, the partition function on the replica space is a summation over $\beta_{1,\dots,2p}, \chi_{1,\dots,2p}$ of
\begin{equation}\label{eq-nega-toSum}
\begin{split}
\frac{1}{k^{4p}}\exp\!\Big[\frac{2\pi i L_{BC}}{k}\big(&(\beta_1 - \beta_3)\chi_2 + (\beta_3 - \beta_5)\chi_4 + (\beta_5 - \beta_1)\chi_6\\
+\,&(\beta_2 - \beta_4)\chi_3 + (\beta_4 - \beta_6)\chi_5 + (\beta_6 - \beta_2)\chi_1\big)\Big]
\end{split}
\end{equation}
with the following $2p = 6$ constraints
\begin{equation}\label{eq-constraint-1}
\begin{split}
1 = \quad&\eta\big[(\beta_1 - \beta_2)L_{AB}\big]\, \eta\big[(\beta_2 - \beta_3)L_{AB}\big]\, \eta\big[(\beta_3 - \beta_4)L_{AB}\big]\\
\times &\eta\big[(\beta_4 - \beta_5)L_{AB}\big]\, \eta\big[(\beta_5 - \beta_6)L_{AB}\big]\, \eta\big[(\beta_6 - \beta_1)L_{AB}\big]\ ,
\end{split}
\end{equation}
The summation of \eqref{eq-nega-toSum} over $\chi_{1,\dots,2p}$ gives $2p$ more constrains
\begin{equation}\label{eq-constraint-2}
\begin{split}
1 = \quad &\eta\big[(\beta_1 - \beta_3)L_{BC}\big]\, \eta\big[(\beta_3 - \beta_5)L_{BC}\big]\, \eta\big[(\beta_5 - \beta_1)L_{BC}\big]\\
\times &\eta\big[(\beta_2 - \beta_4)L_{BC}\big]\, \eta\big[(\beta_4 - \beta_6)L_{BC}\big]\, \eta\big[(\beta_6 - \beta_2)L_{BC}\big]\ .
\end{split}
\end{equation}
Putting \eqref{eq-constraint-2} and \eqref{eq-constraint-1} together and using the property \eqref{eq-property} repeatedly, we end up with
\begin{equation}
\begin{split}
1 = \eta\big[(\beta_1 - \beta_2)L_{AB}\big]\, &\eta\big[\beta_{13}L_{AB}\big]\, \eta\big[\beta_{35}L_{AB}\big]\, \eta\big[\beta_{24}L_{AB}\big]\, \eta\big[\beta_{46}L_{AB}\big]\\
\times&\eta\big[\beta_{13}L_{BC}\big]\, \eta\big[\beta_{35}L_{BC}\big]\, \eta\big[\beta_{24}L_{BC}\big]\, \eta\big[\beta_{46}L_{BC}\big]\ .
\end{split}
\end{equation}
Now we can count the number of solutions. There are 
\begin{enumerate}[\textbullet]
\item $\gcd(k,L_{AB},L_{BC})$ possible $\beta_{13}\in \mathbb{Z}_k$ satisfying $1 = \eta\big[\beta_{13}L_{AB}\big]\, \eta\big[\beta_{13}L_{BC}\big]$. Similar for $\beta_{35}$, $\beta_{24}$ and $\beta_{46}$.
\item $k\cdot \gcd(k,L_{AB})$ possible $\{\beta_{1},\beta_{2}\}\in \mathbb{Z}_k^{\otimes 2}$ satisfying $1 = \eta\big[(\beta_{1} - \beta_{2}) L_{AB}\big]$. 
\end{enumerate}
Therefore, we end up with \eqref{eq-EN-replica2p}
\begin{equation}
\begin{split}
\mathrm{Tr} \big(\rho_{BC}^{\Gamma}\big)^{2p}\big|_{L_{CA = 0}} = \frac{k^{2p}}{k^{4p}}\gcd(k,L_{AB},L_{BC})^{2p -2}\times k \gcd(k,L_{AB})\ ,
\end{split}
\end{equation}
where the factor $k^{2p}$ comes from the summation of $\chi_{1,\dots,2p}$. 
\section{Non-rigorous proof of \eqref{eq-third-negativity}}\label{appx-nega3}
In this appendix we calculate the quantity $\mathrm{Tr} \big(\rho_{BC}^{\Gamma}\big)^{3}$ for the three-link state~\eqref{eq-wave3}. First note that this quantity is permutation symmetric with respect to $A$, $B$, and $C$, since
\begin{equation}\label{eq-rhoBC3}
\mathrm{Tr} \big(\rho_{BC}^{\Gamma}\big)^{3} = \bra{\mathcal{L}^3}^{\otimes 3}\sigma_{A} \sigma_{B} \sigma_{C} \ket{\mathcal{L}^3}^{\otimes 3}
\end{equation}
with $\sigma_{A,B,C}$ the elements of the permutation group
\begin{equation}
\sigma_A = (1)(2)(3)\ ,\quad \sigma_B = (123)\ , \quad \sigma_C = (321)\ .
\end{equation}
The quantity \eqref{eq-rhoBC3} is invariant under renaming of the copies. Multiplying $\sigma_{A,B,C}$ by $(321)$, we obtain
\begin{equation}
(321)\sigma_A = (321) = \sigma_C\ ,\quad (321)\sigma_B = (1)(2)(3) = \sigma_A\ , \quad (321)\sigma_C = (123) = \sigma_B\ .
\end{equation}
thus we can see that \eqref{eq-rhoBC3} is symmetric about $A$, $B$, and $C$. 

Using \eqref{eq-rhoBCT}, we obtain
\begin{equation}
\begin{split}
\mathrm{Tr} \big(\rho_{BC}^{\Gamma}\big)^{3} &=\!\!\! \sum_{\beta_{1,2,3}, \chi_{1,2,3}}\!\frac{1}{k^6}\exp\Big[\frac{2\pi i L_{BC}}{k}\big( \beta_1(\chi_2 - \chi_3) + \beta_2 (\chi_3 - \chi_1) + \beta_3 (\chi_1 - \chi_2)\big)\Big]\\
&\quad\qquad \times \eta\big[\beta_{12}L_{AB}\! -\! \chi_{12}L_{CA}\big]\, \eta\big[\beta_{23}L_{AB}\! -\! \chi_{23}L_{CA}\big]\, \eta\big[\beta_{31}L_{AB}\! -\! \chi_{31}L_{CA}\big]\ ,
\end{split}
\end{equation}
where we use the notation \eqref{eq-notation-ij}. When $L_{BC} = 0$, we have
\begin{equation}
\begin{split}
\mathrm{Tr} \big(\rho_{BC}^{\Gamma}\big)^{3}\big|_{L_{BC = 0}} &=\!\!\! \sum_{\beta_{1,2,3}, \chi_{1,2,3}}\!\frac{1}{k^6}{\color{blue}\eta\big[\beta_{12}L_{AB}\! -\! \chi_{12}L_{CA}\big]}\, {\color{red}\eta\big[\beta_{23}L_{AB}\! -\! \chi_{23}L_{CA}\big]}\\
&= k^{-6}\times k^2\times {\color{blue}k \gcd(k,L_{CA},L_{AB})}\times {\color{red}k \gcd(k,L_{CA},L_{AB})}\\
&= k^{-2}\gcd(k,L_{CA},L_{AB})^2\ ,
\end{split}
\end{equation}
where in the first line we use the property \eqref{eq-property} to remove one non-independent constraint. Since $\mathrm{Tr} \big(\rho_{BC}^{\Gamma}\big)^{3}$ is symmetric about $A,B,C$, we should perform the following improvement
\begin{equation}
\gcd(k,L_{CA},L_{AB}) = \gcd(k,L_{CA},L_{AB},0) \to \gcd(k,L_{CA},L_{AB},L_{BC})\ ,
\end{equation}
thus we end up with \eqref{eq-third-negativity}. We also note here that although we do not have a rigorous derivation of \eqref{eq-third-negativity}, it passes lots of numerical test (we test it numerically for all 3-links with $k\le 18$).  
\section{The Venn diagram}\label{appx-Veen}
An integer can be uniquely understood as the \textit{multiset}\footnote{A multiset generalizes the notion of a set by permitting elements to occur multiple times \cite{enwiki:1298654476}. } of its prime factors. For example, the multiset corresponding to $180$ is $[2,2,3,3,5]$.\footnote{We use the notation $[\cdots]$ to denote a multiset. } In this appendix we use the \textit{Venn diagram}\footnote{The Venn diagram is a diagram style that shows the logical relation between sets~\cite{enwiki:1297058976}.} to illustrate the relationships between integers.\footnote{We thank Jinwei Chu for discussions. } We also present the Venn diagrams of various entanglement measures studied in this paper.
\begin{figure}[htbp]
    \centering
    \includegraphics[scale = 0.95]{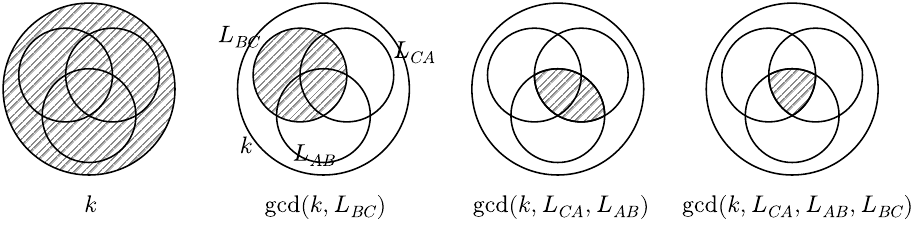}
    \caption{From left to right: The Venn diagram of $k$, $\gcd(k,L_{BC})$, $\gcd(k,L_{CA},L_{AB})$, $\gcd(k,L_{CA},L_{AB},L_{BC})$ respectively. }
    \label{fig-Venn}
\end{figure}

As shown in Fig.~\ref{fig-Venn}, every region represents the multiset corresponding to a nonnegative integer. In this graphical representation, the direct sum of regions corresponds to the multiplication of integers, the direct subtraction of regions corresponds to the division, the intersection of regions corresponds to the gcd function, and the union of regions corresponds to the lcm (least common multiple) function. 

\begin{figure}[htbp]
    \centering
    \includegraphics[scale = 1]{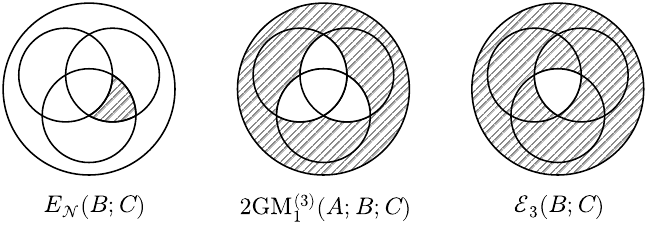}
    \caption{From left to right: The Venn diagram of the argument of the log term in the logarithmic negativity~\eqref{eq-negativity-3link}, the genuine multi-entropy~\eqref{eq-n-GME-3link}, and the third-order negativity~\eqref{eq-third-negativity}. }
    \label{fig-Venn-enatnglement}
\end{figure}

The Venn diagrams are very useful in proving inequalities. For example, for the argument of the logarithmic term in $2\mathrm{GM}_1^{(3)}$ \eqref{eq-n-GME-3link}: 
\begin{equation}
\frac{k\cdot \gcd(k,L_{CA},L_{AB},L_{BC})^2}{\gcd(k,L_{CA},L_{AB}) \gcd(k,L_{AB},L_{BC}) \gcd(k,L_{BC},L_{CA})}\ ,
\end{equation}
its Venn diagram is shown in the middle part of Fig.~\ref{fig-Venn-enatnglement}, from which we can directly see that the inequality \eqref{eq-ineq} holds. 
\section{\boldmath Genuine tri-entropy in (1+1)d Chern-Simons theory\unboldmath}\label{sec-1+1d}
\begin{figure}[htbp]
    \centering
    \includegraphics[scale = 1.1]{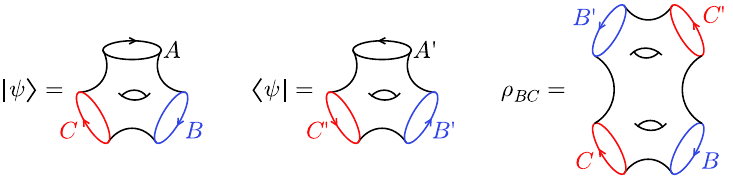}
    \caption{The graphical illustration of $\psi_m^{N_A,N_B,N_C}$ and the unnormalized reduced density matrix $\rho_{BC}$ with $N_A = N_B = N_C = 1$ and $m = 1$. }
    \label{fig-2d}
\end{figure}

In this section we consider the tri-entropy in (1+1)d Chern-Simons theory with finite gauge group $G$. Specifically, we follow the setup in section 3 of \cite{Dwivedi:2020jyx}. The state $\psi^{N_A,N_B,N_C}_m$ is defined on the $N = N_A + N_B + N_C$ numbers of $\mathbb{S}^1$ boundaries of a two-dimensional genus-$m$ manifold, see Fig.~\ref{fig-2d} for an illustration of the $N_A = N_B = N_C$, $m = 1$ case.

Performing the replica trick as in Fig.~\ref{fig-replica-trick}, one can see that the $n^2$-replica geometry is a two-manifold with $g = N n^2 + 2 m n^2 - 2n^2 + 1$ genuses, thus we have
\begin{equation}\label{eq-2d-partition-func}
\mathcal{Z}_{n^2}^{(3)} = \mathcal{Z}\big[\Sigma_{N n^2 + 2 m n^2 - 2 n^2 + 1}\big]\ ,
\end{equation}
and
\begin{equation}
S_{n}^{(3)}(A;B;C) = \frac{1}{1 - n}\frac{1}{n}\log\frac{\mathcal{Z}\big[\Sigma_{N n^2 + 2 m n^2 - 2 n^2 + 1}\big]}{\mathcal{Z}\big[\Sigma_{N + 2m - 1}\big]^{n^2}}\ .
\end{equation}
The partition function \eqref{eq-2d-partition-func} can be expressed in terms of the group-theoretic properties of the finite gauge group $G$ as~\cite{Freed1993}
\begin{equation}
\mathcal{Z}[\Sigma_g] = |G|^{2g - 2}\sum\nolimits_i(\dim R_i)^{2 - 2g}\ ,
\end{equation}
with $R_i$ the irreducible representations of the group $G$ and $\dim R_i$ their dimensions. Therefore, we can obtain
\begin{equation}
S_{1}^{(3)}(A;B;C) = -2 \sum\nolimits_{i}p_i\log p_i\ , \quad \text{with}\ \ p_i = \frac{(\dim R_i)^{4 - 4m - 2N}}{\sum_j (\dim R_j)^{4 - 4m - 2N}}\ .
\end{equation}
The entanglement entropy in this setup is \cite{Dwivedi:2020jyx}
\begin{equation}
S(A;BC) = S(B;CA) = S(C;AB) = - \sum\nolimits_{i}p_i\log p_i\ ,
\end{equation}
thus we end up with
\begin{equation}\label{eq-2d-GM1}
\mathrm{GM}_{1}^{(3)}\big[\psi^{N_A,N_B,N_C}_m\big] = -\frac{1}{2} \sum\nolimits_{i}p_i\log p_i\ .
\end{equation}
We also note that in this setup the logarithmic negativities~\eqref{eq-def-nega} between $A$ and $B$, $B$ and $C$, $C$ and $A$ vanish \cite{Dwivedi:2020jyx},
\begin{equation}
E_{\mathcal{N}}(A;B) = E_{\mathcal{N}}(B;C) = E_{\mathcal{N}}(C;A) = 0\ .
\end{equation}
\section{Consistency between \eqref{eq-n-multiE-3link} and \eqref{eq-Akella}}\label{appx-stabilizerFormula}
A recent paper~\cite{Akella:2025owv} shows that the tripartite multi-entropy for a general tripartite stabilizer state is
\begin{equation}\label{eq-Akella}
S^{(3)}\left(A;B;C\right)=\frac{1}{2}\log\frac{|G|}{|G_{A}||G_{B}||G_{C}|}+\frac{1}{2}\log\frac{|G|}{|G_{AB}\cdot G_{BC} \cdot G_{CA}|}\ ,
\end{equation}
where $G$ denotes the stabilizer group, and $G_A$, $G_B$, $G_C$ are the subgroups that act trivially outside $A$, $B$, $C$ respectively. Similarly, subgroups $G_{AB}$, $G_{BC}$, and $G_{AC}$ act trivially on $C$, $A$, and $B$ respectively. The subgroup $G_{AB}\cdot G_{BC} \cdot G_{CA}$ is defined as
\begin{equation}
    G_{AB}\cdot G_{BC} \cdot G_{CA} = \{ g_1\cdot g_2 \cdot g_3 \mid g_1 \in G_{AB},\ g_2 \in G_{BC},\ g_3 \in G_{CA} \}\ .
\end{equation}
In this appendix, we analytically show that \eqref{eq-Akella} is consistent with~\eqref{eq-n-multiE-3link}. 

The stabilizer group of the 3-link state~\eqref{eq-wave3} is generated by~\cite{Balasubramanian:2018por}
\begin{align}
K_A = X_A Z_B^{L_{AB}} Z_C^{L_{CA}}\ ,\\
K_B = X_B Z_C^{L_{BC}} Z_A^{L_{AB}}\ , \\
K_C = X_C Z_A^{L_{CA}} Z_B^{L_{BC}}\ ,
\end{align}
where $X_A$ is the generalized Pauli $X$ operator acting on subregion $A$, and the generalized Pauli $X$ and $Z$ operators are defined as 
\begin{equation}
X = \sum_{h\in\mathbb{Z}_k} \ket{h + 1\ (\mmod k)}\bra{h}\ ,\quad Z = \sum_{h\in\mathbb{Z}_k}\omega^h\ket{h}\bra{h}\ ,\quad \omega = e^{2\pi i/k}\ .
\end{equation}

The stbilizer group $G$ contains elements of the form $K_A^m K_B^n K_C^p$ with $(m,n,p)\in \mathbb{Z}_k^{\otimes 3}$, thus $|G| = k^3$. The subgroup $G_{BC}$ contains elements of the form $K_B^p K_C^q$ with 
\begin{equation}
p L_{AB} + q L_{CA} \equiv 0\ (\mmod k)\ ,\quad (p,q)\in \mathbb{Z}_k^{\otimes 2}\ ,
\end{equation}
which has $k\gcd(k,L_{CA},L_{AB})$ sets of $(p,q)$ solutions. Similarly we have
\begin{equation}
|G_{AB}| = k\gcd(k,L_{BC},L_{CA})\ ,\quad |G_{BC}| = k\gcd(k,L_{CA},L_{AB})\ ,\quad |G_{CA}| = k\gcd(k,L_{AB},L_{BC})\ .
\end{equation}

The subgroup $G_{A}$ contains elements of the form $K_A^m$ with $m \in \mathbb{Z}_k$ satisfying $m L_{AB} \equiv 0\ (\mmod k)$ and $m L_{CA} \equiv 0\ (\mmod k)$, thus $G_A$ contains $\gcd(k,L_{CA},L_{AB})$ number of elements. Therefore we have
\begin{equation}\label{eq-sizeG}
|G_A| = \gcd(k,L_{CA},L_{AB})\ , \quad |G_B| = \gcd(k,L_{AB},L_{BC})\ ,\quad |G_C| = \gcd(k,L_{BC},L_{CA})\ .
\end{equation}

Now let us finally move to the calculation of $|G_{AB}\cdot G_{BC}\cdot G_{CA}|$. The subgroup ${\color{red}G_{AB}}\cdot G_{BC}\cdot {\color{blue}G_{CA}}$ contains elements of the form 
\begin{equation}\label{eq-GGG}
\begin{split}
&{\color{red}K_{A}^m K_{B}^n} \cdot K_{B}^p K_{C}^q \cdot {\color{blue}K_{C}^r K_{A}^s}\\
= &X_{A}^{m + s} X_{B}^{p + n} X_C^{r + q} Z_A^{(r + q) L_{CA} + (p + n) L_{AB}} Z_B^{(m + s) L_{AB} + (r + q) L_{BC}} Z_C^{(p + n) L_{BC} + (m + s) L_{CA}}\ ,
\end{split}
\end{equation}
with $(m,n,p,q,r,s)\in \mathbb{Z}_k^{\otimes 6}$ and satisfying
\begin{equation}\label{eq-modeq-1}
\begin{cases}
m L_{CA} + n L_{BC} \equiv 0\ (\mmod k)\ ,\\
p L_{AB} + q L_{CA} \equiv 0\ (\mmod k)\ ,\\
r L_{BC} + s L_{AB} \equiv 0\ (\mmod k)\ .
\end{cases}
\end{equation}
Now we need to figure out how many \textit{different} elements are there. We already know $|G_{AB}|$, $|G_{BC}|$ and $|G_{CA}|$. However, when
\begin{equation}\label{eq-modeq-2}
\begin{cases}
m + s \equiv 0\ (\mmod k)\ ,\\
p + n \equiv 0\ (\mmod k)\ ,\\
r + q \equiv 0\ (\mmod k)\ ,
\end{cases}
\end{equation}
Eq. \eqref{eq-GGG} becomes identity, thus we need to divide by the degeneracy,
\begin{equation}
|G_{AB}\cdot G_{BC}\cdot G_{CA}| = \frac{|G_{AB}|\cdot |G_{BC}|\cdot |G_{CA}|}{\# \text{ of solutions of } \eqref{eq-combineMOD}}\ ,
\end{equation}
with \eqref{eq-combineMOD} the combination of \eqref{eq-modeq-1} and \eqref{eq-modeq-2}, 
\begin{equation}\label{eq-combineMOD}
\begin{cases}
m L_{CA} - p L_{BC} \equiv 0\ (\mmod k)\ ,\\
p L_{AB} - r L_{CA} \equiv 0\ (\mmod k)\ ,\\
r L_{BC} - m L_{AB} \equiv 0\ (\mmod k)\ ,
\end{cases}
\!\Longleftrightarrow\ 
\begin{pmatrix}
\mB & -\mA & \\
 & \mC & -\mB\\
-\mC &  & \mA
\end{pmatrix}\begin{pmatrix}
m\\
p\\
r
\end{pmatrix} \equiv\mathbf{0}\ (\mmod k)\ ,
\end{equation}
where we introduce $\mA := L_{BC}$, $\mB := L_{CA}$ and $\mC := {L_{AB}}$.

The number of solutions of a system of congruences
\begin{equation}
    M \mathbf{x} \equiv \mathbf{0}\ (\mmod k)\ ,\quad \mathbf{x} \in \mathbb{Z}_k^\mathrm{n}
\end{equation}
can be determined using the Smith normal form (SNF)~\cite{enwiki:1288152750} of the corresponding coefficient matrix $M$. The main idea is to find two nonsingular matrices $U$ and $V$ such that
\begin{equation}
    UMV=D=\mathrm{diag}(d_1, d_2, d_3, \cdots, d_\mathrm{r}, 0, \cdots, 0)\ ,
\end{equation}
where the diagonal elements $d_i$ satisfy $d_i|d_{i+1}$, and $\mathrm{r}=\mathrm{rank}(M)$. Using this method, one can transform the original system of congruences to
\begin{equation}
    D\mathbf{y} \equiv \mathbf{0}\ (\mmod k)\ , \quad \mathbf{y} \in \mathbb{Z}_k^\mathrm{n}\ ,
\end{equation}
with $\mathbf{y}=V^{-1}\mathbf{x}$.
The diagonal element $d_i$ can be computed as
\begin{equation}\label{eq-d-m}
    d_i = \frac{m_i(M)}{m_{i-1}(M)}\ , 
\end{equation}
where $m_i(M)$ equals the greatest common divisor of the determinants of all $i\times i$ minors of the matrix $M$ and $m_0(M)=1$. Then the number of solutions is
\begin{equation}\label{eq-NumSol}
k^{(\mathrm{n} - \mathrm{r})} \prod^{\mathrm{r}}_{i = 1} \mathrm{gcd}(k,d_i) \ .
\end{equation}
In our case, $\mathrm{n} = 3$, $\mathrm{r} = 2$, and it can be easily computed that
\begin{equation}\label{eq-m1m2}
\begin{cases}
    m_1=\mathrm{gcd}(\mA, \mB, \mC)\ , \\
    m_2= \gcd(\mA^2, \mB^2, \mC^2, \mA\mB, \mB\mC, \mC\mA)\ .
\end{cases}
\end{equation}
The $m_2$ in \eqref{eq-m1m2} can be further simplified as
\begin{equation}\label{eq-gcdeq}
\begin{split}
m_2 &= \gcd(\mA^2, \mB^2, \mC^2, \mA\mB, \mB\mC, \mC\mA)\\
&= \gcd(\mA, \mB, \mC)^2 \gcd(\ma^2, \mb^2, \mc^2, \ma\mb, \mb\mc, \mc\ma)\\
&= \gcd(\mA, \mB, \mC)^2 \gcd(\gcd(\ma^2, \ma\mb, \ma\mc), \gcd(\mb\ma, \mb^2, \mb\mc), \gcd(\mc\ma, \mc\mb, \mc^2))\\
&= \gcd(\mA, \mB, \mC)^2 \gcd(\ma\gcd(\ma, \mb, \mc), \mb\gcd(\ma, \mb, \mc), \mc\gcd(\ma, \mb, \mc))\\
&= \gcd(\mA, \mB, \mC)^2\ ,
\end{split}
\end{equation}
where in the second line we introduce
\begin{equation}
\ma = \frac{\mA}{\gcd(\mA, \mB, \mC)}\ ,\quad \mb = \frac{\mB}{\gcd(\mA, \mB, \mC)}\ ,\quad \mc = \frac{\mC}{\gcd(\mA, \mB, \mC)}\ ,
\end{equation}
and thus we have $\gcd(\ma, \mb, \mc) = 1$, which is applied to the fourth line of \eqref{eq-gcdeq}. 

With the results of $m_1$ and $m_2$, according to \eqref{eq-d-m} and \eqref{eq-NumSol} we have 
\begin{equation}
\# \text{ of solutions of } \eqref{eq-combineMOD} = k \gcd(k, \gcd(L_{BC}, L_{CA}, L_{AB}))^2\ ,
\end{equation}
and thus
\begin{equation}\label{eq-sizeGGG}
|G_{AB}\cdot G_{BC}\cdot G_{CA}| = \frac{k\gcd(k,L_{BC},L_{CA})\cdot k\gcd(k,L_{CA},L_{AB})\cdot k\gcd(k,L_{AB},L_{BC})}{k \gcd(k, L_{BC}, L_{CA}, L_{AB})^2}\ .
\end{equation}
Inserting $|G| = k^3$, \eqref{eq-sizeG} and \eqref{eq-sizeGGG} into \eqref{eq-Akella}, we obtain explicitly \eqref{eq-n-multiE-3link} with $n = 1$. 



\bibliographystyle{JHEP}
\bibliography{biblio}
\end{document}